\documentclass[twocolumn,floatfix,aps,superscriptaddress,eqsecnum,prb]{revtex4}
\usepackage{graphicx}
\usepackage{amsmath}
\usepackage{amssymb}
\usepackage{bm}

\usepackage{color}

\begin{document}

\title{Chirality blockade of Andreev reflection in a magnetic Weyl semimetal}
\author{N. Bovenzi}
\affiliation{Instituut-Lorentz, Universiteit Leiden, P.O. Box 9506, 2300 RA Leiden, The Netherlands}
\author{M. Breitkreiz}
\affiliation{Instituut-Lorentz, Universiteit Leiden, P.O. Box 9506, 2300 RA Leiden, The Netherlands}
\author{P. Baireuther}
\affiliation{Instituut-Lorentz, Universiteit Leiden, P.O. Box 9506, 2300 RA Leiden, The Netherlands}
\author{T. E. O'Brien}
\affiliation{Instituut-Lorentz, Universiteit Leiden, P.O. Box 9506, 2300 RA Leiden, The Netherlands}
\author{J. Tworzyd{\l}o}
\affiliation{Institute of Theoretical Physics, Faculty of Physics, University of Warsaw, ul.\ Pasteura 5, 02--093 Warszawa, Poland}
\author{\.{I}. Adagideli}
\affiliation{Faculty of Engineering and Natural Sciences, Sabanci University, Orhanli-Tuzla, 34956 Istanbul, Turkey}
\author{C. W. J. Beenakker}
\affiliation{Instituut-Lorentz, Universiteit Leiden, P.O. Box 9506, 2300 RA Leiden, The Netherlands}

\date{April 2017}
\begin{abstract}
A Weyl semimetal with broken time-reversal symmetry has a minimum of two species of Weyl fermions, distinguished by their opposite chirality, in a pair of Weyl cones at opposite momenta $\pm K$ that are displaced in the direction of the magnetization. Andreev reflection at the interface between a Weyl semimetal in the normal state (N) and a superconductor (S) that pairs $\pm K$ must involve a switch of chirality, otherwise it is blocked. We show that this ``chirality blockade'' suppresses the superconducting proximity effect when the magnetization lies in the plane of the NS interface. A Zeeman field at the interface can provide the necessary chirality switch and activate Andreev reflection.
\end{abstract}
\maketitle

\section{Introduction}

Spin-momentum locking is a key feature of topological states of matter: In both topological insulators and topological semimetals the massless quasiparticles are governed by a Hamiltonian $H_\pm=\pm v_{\rm F}\bm{p}\cdot\bm{\sigma}$ that ties the direction of motion to the spin polarization.\cite{Has10,Qi11,Has17,Yan17} In a topological insulator the $\pm$ sign distinguishes spatially separated states, e.g., the opposite edges of a quantum spin-Hall insulator along which a spin-up electron moves in opposite directions.\cite{Kon08} In a topological semimetal the $\pm$ sign distinguishes Weyl cones in the band structure. A magnetic Weyl semimetal has the minimum number of two Weyl cones centered at opposite points $\pm\bm{K}$ in the Brillouin zone, containing left-handed and right-handed Weyl fermions displaced in the direction of the magnetization.\cite{Bur11}

\begin{figure}[tb]
\centerline{\includegraphics[width=0.9\linewidth]{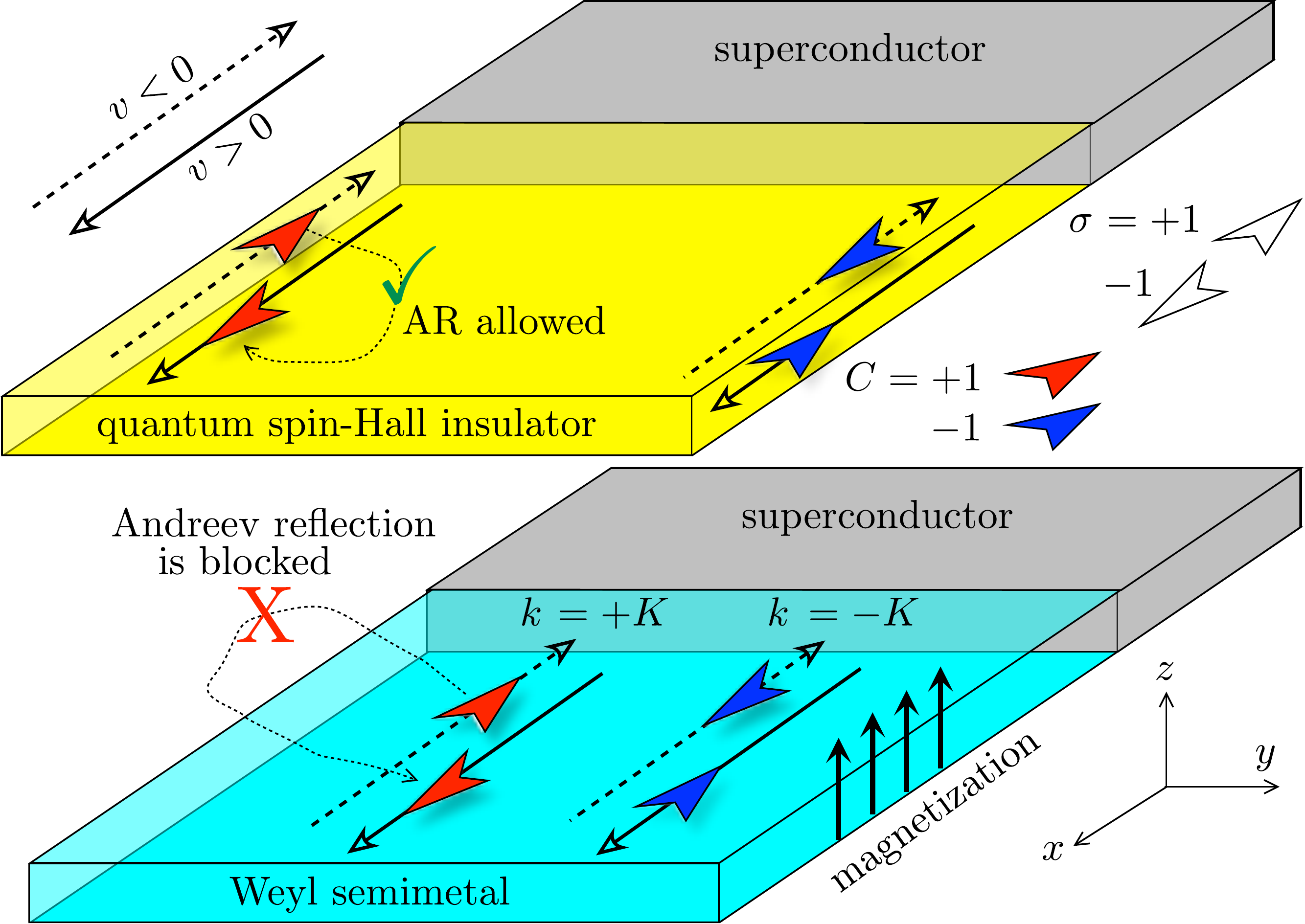}}
\caption{Andreev reflection (AR) from a superconductor in a quantum spin-Hall insulator (top panel) and in a Weyl semimetal (bottom panel.) The red and blue wedges designate electron and hole quasiparticles (Weyl fermions) moving towards or away from the interface (solid versus dashed arrows indicate $v$ in the $\pm x$ direction). The orientation of the wedge distinguishes the polarization $\sigma=\pm 1$ of the spin band and the color indicates the chirality $C={\rm sign}\,(v\sigma)$. Andreev reflection switches $\sigma$ and $v$, which is blocked if it must also switch $C$. 
}
\label{fig_layout}
\end{figure}

It is the purpose of this paper to point out that the switch in chirality between the Weyl cones forms an obstacle to Andreev reflection from a superconductor with conventional, spin-singlet \textit{s}-wave pairing, when the magnetization lies in the plane of the normal-superconductor (NS) interface. The obstruction is illustrated in Fig.\ \ref{fig_layout}. Andreev reflection is the backscattering of an electron as a hole, accompanied by the transfer of a Cooper pair to the superconductor. For a given spin band and a given Weyl cone, electrons and holes move in the same direction,\cite{note0} so backscattering must involve either a switch in spin band ($\bm{\sigma}\mapsto -\bm{\sigma}$) or a switch in Weyl cone ($\bm{K}\mapsto -\bm{K}$), but not both. This is at odds with the requirement that zero spin and zero momentum is transferred to the Cooper pair.

This ``chirality blockade'' of Andreev reflection is specific for the conical dispersion in a Weyl semimetal, and it does not appear in other contexts where spin-momentum locking plays a role. In a quantum spin-Hall insulator, there is no need to switch the chirality because the hole can be reflected along the same edge as the incident electron.\cite{Har14} There is a formal similarity with graphene,\cite{Bee09} where Andreev reflection switches between valleys at $\pm\bm{K}$, but there $\bm{\sigma}$ is an orbital pseudospin and the real spin is not tied to the direction of motion.

We will show that the chirality blockade can be lifted by breaking the requirement of zero-spin transfer with a Zeeman field. We also discuss the subtle role played by inversion symmetry, by contrasting a \textit{scalar} with a \textit{pseudoscalar} pair potential.\cite{Far16} The absence of the chirality blockade for pseudoscalar pairing explains why it did not appear in the many previous studies of Andreev reflection in a Weyl semimetal.\cite{Uch14,Che13,Kha16,Mad17,Kim16,Sal16,Kha17,Agg17,Wan17}

The outline of this paper is as follows. In the next section, we introduce the model of an NS junction between a Weyl semimetal and a conventional superconductor. The $8\times 8$ Bogoliubov-De Gennes Hamiltonian is block-diagonalized in Sec.\ \ref{blockdiagonalization}, after which the chirality blockade of Andreev reflection is obtained in Sec.\ \ref{sec_AR}. In the next section \ref{ARactivation}, we show how to remove the blockade by a spin-active interface or by an inversion-symmetry breaking interface. As an experimental signature, we calculate the conductance of the NS junction in Sec.\ \ref{sec_cond}. To eliminate the effects of a lattice mismatch, we consider in Sec.\ \ref{WeylWeylsec} the NS junction between a Weyl semimetal and a Weyl superconductor --- which shows the same chirality blockade for a scalar spin-singlet pair potential. More general pairing symmetries (spin-triplet and pseudoscalar spin-singlet) are considered in an Appendix. The Josephson effect in an SNS junction is studied in Sec.\ \ref{FermiJosephson}. We conclude in Sec.\ \ref{sec_discuss}. 

\section{Model of a Weyl semimetal -- conventional superconductor junction}
\label{sec_model}

We study the junction between a Weyl semimetal in the normal state (N) and a conventional (spin-singlet, \textit{s}-wave) superconductor (S), by first considering separately the Hamiltonians in the two regions and then modeling the interface. 

Throughout the paper, we take the configuration of Fig.\ \ref{fig_layout} (bottom panel), with the magnetization along $z$ in the plane of the NS interface at $x=0$. An out-of-plane rotation of the magnetization by an angle $\alpha$ does not change the results for isotropic Weyl cones, provided that the Fermi surfaces of opposite chirality are not coupled upon reflection at the interface. The geometric condition for this is $\cos \alpha\geq k_{\rm F}/K$, with $k_{\rm F}$ the Fermi wave vector and $(0,0,\pm K)$ the location of the two Weyl points. We assume $k_{\rm F}/K\ll 1$ in order to have well-resolved Weyl cones, and then there is a broad range of magnetization angles $\alpha$ over which our analysis applies.

\subsection{Weyl semimetal region}

The Weyl semimetal in the region $x>0$ has the generic Hamiltonian\cite{Yan11,Cho11,Vaz13}
\begin{subequations}
\label{HW}
\begin{align}
H_{\rm W}(\bm{k})={}&\tau_z(\sigma_x t_x\sin k_x+\sigma_y t_y\sin k_y+\sigma_z t_z\sin k_z)\nonumber\\
&+m_{\bm{k}}\tau_x\sigma_0+\beta\tau_0\sigma_z-\mu_{\rm W}\tau_0\sigma_0,\label{HWa}\\
m_{\bm{k}}={}&m_0+t'_x(1-\cos k_x)+t'_y(1-\cos k_y)\nonumber\\
&+t'_z(1-\cos k_z).\label{HWb}
\end{align}
\end{subequations}
The units are normalized by $\hbar\equiv 1$ and lattice constant $a_0\equiv 1$. The Pauli matrices $\tau_\alpha$ and $\sigma_\alpha$ refer to orbital and spin degrees of freedom (with $\tau_0$, $\sigma_0$ the $2\times 2$ unit matrix). The Weyl points are at $\bm{k}=(0,0,\pm K)$, with
\begin{equation}
K^2\approx\frac{{\beta^2-m_0^2}}{{ t_z^2+t'_z m_0}}\label{Kapprox}
\end{equation}
displaced by the magnetization $\beta$ in the $z$-direction. The mass term $m_{\bm{k}}$ ensures that there are no other states near the Fermi energy, so that we have the minimal number of two Weyl cones of opposite chirality.

While time-reversal symmetry is broken by the magnetization $\beta$, the inversion symmetry of the material is preserved:
\begin{equation}
\tau_x H_{\rm W}(-\bm{k})\tau_x=H_{\rm W}(\bm{k}).\label{inversionsymmetryW}
\end{equation}
The presence of inversion symmetry plays a crucial role when superconductivity enters, because the pair potential couples electrons and holes at opposite momentum.

To describe the superconducting proximity effect we add the electron-hole degree of freedom $\nu$, with electron and hole Hamiltonians related by the operation of time-reversal:
\begin{equation}
H_{\rm W}^{(e)}(\bm{k})=H_{\rm W}(\bm{k}),\;\;H_{\rm W}^{(h)}(\bm{k})=\sigma_yH^\ast_{\rm W}(-\bm{k})\sigma_y.\label{Hehrelation}
\end{equation}  
The two Hamiltonians are incorporated in the Bogoliubov-De Gennes (BdG) Hamiltonian
\begin{align}
{\cal H}_{\rm W}={}&\begin{pmatrix}
H_{\rm W}^{(e)}&0\\
0&-H_{\rm W}^{(h)}
\end{pmatrix}\nonumber\\
={}&\nu_z\tau_z(\sigma_x t_x\sin k_x+\sigma_y t_y\sin k_y+\sigma_z t_z\sin k_z)\nonumber\\
&+m_{\bm{k}}\nu_z\tau_x\sigma_0+\beta\nu_0\tau_0\sigma_z-\mu_{\rm W}\nu_z\tau_0\sigma_0.\label{HBdGW}
\end{align}
Electron-hole symmetry is expressed by
\begin{equation}
\nu_y\sigma_y{\cal H}^\ast_{\rm W}(-\bm{k})\nu_y\sigma_y=-{\cal H}_{\rm W}(\bm{k}).\label{ehsymmetryW}
\end{equation}
Note that the electron-hole symmetry operation squares to $+1$, as it should in symmetry class D (fermions without spin-rotation or time-reversal symmetry). 

\subsection{Superconducting region}

The region $x<0$ contains a conventional spin-singlet \textit{s}-wave superconductor (real pair potential $\Delta_0$), with BdG Hamiltonian
\begin{equation}
{\cal H}_{\rm S}=\begin{pmatrix}
p^2/2m-\mu_{\rm S}&\Delta_0\\
\Delta_0&-p^2/2m+\mu_{\rm S}
\end{pmatrix}.\label{HSdef}
\end{equation}

For a chemical potential $\mu_{\rm S}\gg \mu_{\rm W}$, the momentum components $p_y,p_z$ parallel to the NS interface at $x=0$ can be neglected relative to the perpendicular component $p_x$. We expand $p_x=\pm p_{\rm F}+k_x$ around the Fermi momentum $p_{\rm F}=mv_{\rm F}$ (with $\mu_{\rm S}=p_{\rm F}^2/2m$), by carrying out the unitary transformation
\begin{align}
{\cal H}_{\rm S}&\mapsto e^{-i\tau_z p_{\rm F}x}{\cal H}_{\rm S}e^{i\tau_z p_{\rm F}x}\nonumber\\
&=v_{\rm F}k_x\nu_z\tau_z\sigma_0+\Delta_0\nu_x\tau_0\sigma_0+{\cal O}(k_x^2).\label{HSlinearized}
\end{align}
Left-movers and right-movers in the $x$-direction are distinguished by the $\tau$ degree of freedom, and we have inserted a $\sigma_0$ Pauli matrix to account for the spin degeneracy in S. 

Electron-hole symmetry in S is expressed by
\begin{equation}
\nu_y\tau_x\sigma_y{\cal H}^\ast_{\rm S}(-k_x)\nu_y\tau_x\sigma_y=-{\cal H}_{\rm S}(k_x).\label{ehsymmetryS}
\end{equation}
There is an additional $\tau_x$ Pauli matrix, in comparison with the corresponding symmetry relation \eqref{ehsymmetryW} in N, to account for the switch from $+p_{\rm F}$ to $-p_{\rm F}$. (The electron-hole symmetry operation still squares to $+1$.)

\subsection{Interface transfer matrix}

The wave functions $\psi_{\rm W}$ and $\psi_{\rm S}$ on the two sides of the NS interface at $x=0$ are related by a transfer matrix,
\begin{equation}
\psi_{\rm S}=(t_x/v_{\rm F})^{1/2}{\cal M}\psi_{\rm W},\;\;{\cal M}=\begin{pmatrix}
M_{e}&0\\
0&M_{h}
\end{pmatrix},\label{MSWdef}
\end{equation}
which ensures that particle current is conserved across the interface. We assume that the interface does not couple electrons and holes,\cite{note1} hence the block-diagonal structure, and we also assume that ${\cal M}$ is independent of energy. The symmetry relations \eqref{ehsymmetryW} and \eqref{ehsymmetryS} imply that the electron and hole transfer matrices are related by
\begin{equation}
M_{h}=\tau_x\sigma_y M_{e}^\ast\tau_0\sigma_y.\label{MeMhrelation}
\end{equation}

Particle current conservation is expressed by
\begin{equation}
\langle\psi_{\rm S}|v_{\rm F}\nu_z\tau_z\sigma_0|\psi_{\rm S}\rangle=
\langle\psi_{\rm W}|t_x\nu_z\tau_z\sigma_x|\psi_{\rm W}\rangle,\label{currentNS}
\end{equation}
where we have also linearized $H_{\rm W}$ in $k_x$. The resulting restriction on the electron transfer matrix is
\begin{equation}
M_{e}^\dagger \tau_z\sigma_0 M_{e}=\tau_z\sigma_x.\label{currentconservation}
\end{equation}
Eq.\ \eqref{MeMhrelation} then implies that the hole transfer matrix $M_h$ satisfies the same restriction.

It is helpful to factor out the unitary matrix $\Xi_{0}$,
\begin{equation}
M_{e}\equiv \Xi \Xi_{0},\;\;\Xi_{0}=\exp\left[i\frac{\pi}{4}\tau_x(\sigma_0-\sigma_x)\right],\label{M0Xi}
\end{equation}
with $\Xi_{0}\tau_z \Xi_{0}^\dagger \tau_z =\Xi_{0}^2=\sigma_x$, because now instead of Eq.\ \eqref{currentconservation} we have a quasi-unitarity restriction
\begin{equation}
\Xi^{-1}=\tau_z\Xi^\dagger\tau_z\label{Xiconstraint}
\end{equation}
that is satisfied by the unit matrix.

The corresponding factorization of the hole transfer matrix is
\begin{equation}
M_h=\tau_x\sigma_y (\Xi \Xi_{0})^\ast\tau_0\sigma_y,\label{MhM0}
\end{equation}
as required by the electron-hole symmetry \eqref{MeMhrelation}. For later use we give the inverse
\begin{equation}
M_h^{-1}=(\sigma_y \Xi_{0}\sigma_y)(\tau_z\sigma_y\Xi^{\rm T}\tau_z\sigma_y)\tau_x,\label{MhinversionM0}
\end{equation}
in view of the quasi-unitarity \eqref{Xiconstraint}. (The superscript T denotes the transpose of a matrix.)

As an aside, we note that if the interface preserves time-reversal symmetry, we have the additional restriction
\begin{equation}
\Xi=\tau_x\sigma_y\Xi^{\ast}\tau_x\sigma_y.
\label{XiTRS}
\end{equation}
Inversion symmetry is expressed by
\begin{equation}
\Xi=\tau_x\Xi^{-1}\tau_x.\label{Xiinversion}
\end{equation}

\section{Block-diagonalization of the Weyl Hamiltonian}
\label{blockdiagonalization}

For the mode-matching calculations at the NS interface it is convenient to block-diagonalize ${\cal H}_{\rm W}$ in the $\tau$ degree of freedom, by means of the unitary transformation\cite{Bai17}
\begin{equation}
\begin{split}
&\tilde{\cal H}_{\rm W}={\cal U}{\cal H}_{\rm W}{\cal U}^\dagger,
\;\;{\cal U}=\begin{pmatrix}
i\tau_y\sigma_z\Omega_{\theta}&0\\
0&\Omega_{\theta}
\end{pmatrix},\\
&\Omega_\theta=\exp(-\tfrac{1}{2}i\theta\tau_y\sigma_z),
\end{split}\label{Omegadef}
\end{equation}
with a $\bm{k}$-dependent angle $\theta\in(0,\pi)$ defined by
\begin{equation}
\begin{split}
&\cos\theta=-\frac{t_z\sin k_z}{M_{\bm{k}}},\;\;\sin\theta=\frac{m_{\bm{k}}}{M_{\bm{k}}},\\
&M_{\bm{k}}=\sqrt{m_{\bm{k}}^2+t_z^2\sin^2 k_z}.
\end{split}\label{phidef}
\end{equation}
Note that ${\cal U}$ satisfies
\begin{equation}
{\cal U}(\bm{k})=\nu_y\sigma_y{\cal U}^\ast(-\bm{k})\nu_y\sigma_y,\label{Omegathetasymmetry}
\end{equation}
because $\bm{k}\mapsto -\bm{k}$ maps $\theta\mapsto\pi-\theta$, so the electron-hole symmetry relation \eqref{ehsymmetryW} for ${\cal H}_{\rm W}$ is preserved upon the unitary transformation. 

The transformed Hamiltonian,
\begin{align}
\tilde{\cal H}_{\rm W}(\bm{k})&=\nu_z\tau_z(\sigma_x t_x\sin k_x+\sigma_y t_y\sin k_y)+M_{\bm{k}}\nu_0\tau_z\sigma_z\nonumber\\
&+\beta\nu_0\tau_0\sigma_z-\mu_{\rm W}\nu_z\tau_0\sigma_0,\label{Htransformed}
\end{align}
is block-diagonal in $\tau$.  The Weyl cones are in the $\tau=-1$ block, which has low-energy states near $\bm{k}=(0,0,\pm\beta/t_z)$ when $M_{\bm k}\approx\beta$. The $\tau=+1$ block is pushed to higher energies of order $2\beta$. 

The unitary transformation changes the wave function in N as $\tilde{\psi}_{\rm W}={\cal U}\psi_{\rm W}$, and hence the matching equation \eqref{MSWdef} becomes 
\begin{equation}
\psi_{\rm S}=(t_x/v_{\rm F})^{1/2}{\cal M}{\cal U}^\dagger\tilde{\psi}_{\rm W}.\label{MSWdeftransformed}
\end{equation}

\section{Andreev reflection}
\label{sec_AR}

At excitation energies $E$ below the superconducting gap $\Delta_0$, an electron incident on the superconductor from the Weyl semimetal is reflected, either as an electron (normal reflection, with amplitude $r_{ee}$) or as a hole (Andreev reflection, with amplitude $r_{he}$). We calculate these reflection amplitudes, initially restricting ourselves to normal incidence on the NS interface, in order to simplify the formulas. The angular dependence is included in Sec.\ \ref{sec_cond}, when we calculate the conductance. 

We include the energy dependence of the reflection amplitudes, but since we assume only the low-energy states in the $\tau=-1$ block are propagating our analysis is restricted to $|E|\lesssim\beta$. Typically $\beta\simeq 100\,{\rm meV}$ is much larger than $\Delta_0\simeq 0.1\,{\rm meV}$, so this covers the relevant energy range.

\subsection{Effective boundary condition at the NS interface}
\label{sec_boundarycond}

As in the analogous problem for graphene,\cite{Tit06} the effect of the superconducting region $x<0$ on the Weyl semimetal region $x>0$ can be described by an effective boundary condition on the wave functions in the limit $x\rightarrow 0$ from above, indicated as $x=0^+$.

According to the Hamiltonian \eqref{HSlinearized}, the propagation of the wave function into the superconductor at energy $E$ is governed by the differential equation
\begin{equation}
v_{\rm F}\frac{\partial}{\partial x}\psi(x)=\bigl(iE\nu_z+\Delta_0\nu_y\bigr)\tau_z\sigma_0\psi(x)\equiv X_{\rm S}\psi(x).
\end{equation}
The eigenvalues of $X_{\rm S}$ are $\pm\sqrt{\Delta_0^2-E^2}$. To ensure a decaying wave function in the S region $x<0$ for $|E|<\Delta_0$, the state $\psi_{\rm S}$ at $x=0^-$ should be a linear superposition of the four eigenvectors with positive eigenvalue. This is expressed by the boundary condition
\begin{equation}
\begin{split}
&\nu_x\psi_{\rm S}=\exp(i\alpha\nu_z\tau_z\sigma_0)\psi_{\rm S},\\
&\alpha={\rm arccos}(E/\Delta_0)\in(0,\pi/2).
\end{split}\label{psiSboundary}
\end{equation}

If we decompose $\psi_{\rm S}=(\psi_e,\psi_h)$ into electron and hole components, the boundary condition can be written as
\begin{equation}
\psi_h(0^-)=\exp(i\alpha \tau_z)\psi_e(0^-).\label{psihpsierelation}
\end{equation}
This is a special case of the more general relation between electron and hole wave functions at an NS interface derived in App.\ \ref{app_boundarycond}.

The combination of Eqs.\  \eqref{MSWdeftransformed} and \eqref{psihpsierelation} gives on the Weyl semimetal side of the NS interface the relation
\begin{equation}
\begin{split}
&\tilde{\psi}_h(0^+)= {\cal T}\tilde{\psi}_e(0^+),\\
&{\cal T}=-i\Omega_\theta M_h^{-1}\exp(i\alpha \tau_z)M_e\Omega^\dagger_\theta\tau_y\sigma_z,
\end{split}\label{curlyTdef}
\end{equation}
which can be worked out as
\begin{subequations}
\label{curlyTresult}
\begin{align}
{\cal T}={}&-i\Omega_\theta(\sigma_y \Xi_{0}\sigma_y)(\tau_z\sigma_y\Xi^{\rm T}\tau_z\sigma_y)\tau_x\exp(i\alpha \tau_z)\Xi\nonumber\\
&\hspace*{4cm}\cdot \Xi_{0}\Omega^\dagger_\theta\tau_y\sigma_z\nonumber\\
={}&U_\theta^\dagger \tau_x\sigma_x\Xi^{\rm T}\tau_y\sigma_y\exp(i\alpha \tau_z)\Xi U_\theta,\label{curlyTresulta}\\
U_\theta\equiv{}&\Xi_{0}\Omega^\dagger_\theta\tau_y\sigma_z,\label{Uthetadef}
\end{align}
\end{subequations}
upon substitution of Eq.\ \eqref{MhinversionM0} and using $\tau_y\sigma_z(\sigma_y \Xi_{0}\sigma_y)\tau_y\sigma_z=\Xi_{0}^\dagger$.

\subsection{Reflection amplitudes}
\label{sec_reflectionampl}

We consider an incident mode $\psi_{\rm incident}=(\psi_{e,{\rm inc}},\psi_{h,{\rm inc}})$ without a hole component, $\psi_{h,{\rm inc}}=0$, and initially take the simplest case of normal incidence, when $k_y=0$ and $k_z=\pm K$ is at one of the two Weyl points. (The dependence on the angle of incidence is included later on.) We work in the transformed basis from Section \ref{blockdiagonalization}, when both Weyl points are in the $\tau_z=-1$ band.

The incident electron wave function $\tilde{\psi}_{e,{\rm inc}}=(0,0,1,1)$ has $\sigma_x=+1$ in the $\tau_z=-1$ band, so that its velocity $t_x\nu_z\tau_z\sigma_x$ is in the negative $x$-direction. The reflected wave function $\tilde{\psi}_{\rm reflected}=(\tilde{\psi}_{e,{\rm refl}},\tilde{\psi}_{h,{\rm refl}})$ contains an electron component $\tilde{\psi}_{e,{\rm refl}}=r_{ee}(0,0,1,-1)$ with $\sigma_x=-1$, and a hole component $\tilde{\psi}_{h,{\rm refl}}=r_{he}(0,0,1,1)$ with $\sigma_x=+1$, both waves propagating in the positive $x$-direction. The reflected waves are related to the incident wave by the normal reflection amplitude $r_{ee}$ and the Andreev reflection amplitude $r_{he}$.

At the interface the propagating modes in the $\tau_z=-1$ band may excite evanescent modes in the $\tau_z=+1$ band. Their wave function $\tilde{\psi}_{\rm evan}$ in N is an eigenstate of $\nu_z\sigma_y$ with eigenvalue $+1$, so that the Hamiltonian \eqref{Htransformed} produces a decay for $x\rightarrow\infty$. The electron and hole components of the evanescent mode are $\tilde{\psi}_{e,{\rm evan}}=a(1,i,0,0)$ and $\tilde{\psi}_{h,{\rm surf}}=b(1,-i,0,0)$, with unknown amplitudes $a,b$.

The boundary condition \eqref{curlyTdef} then equates the vectors
\begin{equation}
\begin{pmatrix}
b\\
-ib\\
r_{he}\\
r_{he}
\end{pmatrix}=
{\cal T}
\begin{pmatrix}
a\\
ia\\
1+r_{ee}\\
1-r_{ee}
\end{pmatrix}.\label{reflection_equations}
\end{equation}
There is no dependence on the chemical potential $\mu_{\rm W}$ in the Weyl semimetal for normal incidence.

For an inactive interface, with $\Xi=1$, we have
\begin{equation}
{\cal T}=\tau_y\sigma_z\cos\alpha-i\tau_x\sigma_y\sin\alpha,\label{TXiis1}
\end{equation}
and we find
\begin{equation}
r_{ee}=-ie^{-2i\alpha},\;\;r_{he}=0,\label{reerheTxiis1}
\end{equation}
i.e. fully suppressed Andreev reflection at all energies (and also at all angles of incidence, see Sec.\ \ref{sec_cond}). For $E<\Delta_0$ the incident electron is reflected as an electron with unit probability, without any transfer of a Cooper pair into the superconductor. For $E>\Delta_0$ the angle $\alpha=-i\,{\rm arcosh}\,(E/\Delta_0)$ is imaginary and the incident electron is partly transmitted through the NS interface --- but still without any Cooper pair transfer.

\section{Activation of Andreev reflection}
\label{ARactivation}

Andreev reflection can be restored by a suitably chosen interface potential. We examine two types of interfaces, one that breaks time-reversal symmetry by a Zeeman coupling to the spin, and another that breaks inversion symmetry by a  tunnel coupling to the orbital degree of freedom. 

\subsection{Spin-active interface}
\label{activate_spin}

We consider an interface with a Zeeman Hamiltonian $H_{\rm interf}=g\mu_{\rm B}\bm{B}\cdot\bm{\sigma}$ on the S side, which gives a transfer matrix
\begin{equation}
\Xi=\exp\bigl[i(\ell/v_{\rm F})\tau_z H_{\rm interf}\bigr]=\exp\bigl[i\gamma\tau_z(\bm{n}\cdot\bm{\sigma})\bigr],\label{Xibeta}
\end{equation}
with $\gamma=g\mu_{\rm B}B\ell/v_{\rm F}$, $\bm{n}$ a unit vector in the $\bm{B}$-direction, and $\ell$ is the thickness of the interface layer. The superconducting coherence length $\xi=\hbar v_{\rm F}/\Delta_0$ is an upper bound on $\ell$, and hence $\gamma\lesssim E_{\rm Zeeman}/\Delta_0$, with $E_{\rm Zeeman}=g\mu_{\rm B}B$ the Zeeman spin splitting.

Depending on the direction of the field, we find the Andreev reflection amplitudes
\begin{subequations}
\label{rhespinactive}
\begin{align}
&H_{\rm interf}=B_x\sigma_x\Rightarrow
r_{he}=-\frac{ 2\cos\alpha \sin 2 \gamma \sin \theta}{ \sin^2 2 \gamma \sin^2 \theta+e^{2i\alpha}},\\
&H_{\rm interf}=B_y\sigma_y\Rightarrow
r_{he}=\frac{2 i \sin \alpha \sin 2 \gamma \cos \theta}{ \sin^2 2 \gamma \cos^2  \theta-e^{2i\alpha}},\\
&H_{\rm interf}=B_z\sigma_z\Rightarrow
r_{he}=-\frac{2i\cos\alpha\sin 2\gamma }{\sin^2 2\gamma+e^{2i\alpha}}.
\end{align}
\end{subequations}
At the Fermi level ($E=0\Rightarrow\alpha=\pi/2$), we have $r_{he}=0$ for $\bm{B}$ in the $x$-direction or in the $z$-direction, while a field in $y$-direction activates the Andreev reflection. 

For $m_0\ll\beta\ll t_z$ we may approximate $K\approx\beta/t_z\ll 1$, $\sin\theta\approx \beta/2t_z\ll 1$ and $\cos\theta\approx \mp 1$. The Andreev reflection probability $R_{he}=|r_{he}|^2$ at the Fermi level for $\bm{B}$ in the $y$-direction is then given by
\begin{equation}
R_{he}=\frac{4\sin^2 2\gamma}{(1+\sin^2 2\gamma)^2}.\label{Rheresult}
\end{equation}
It oscillates with $\gamma$, reaching a maximum of unity when $\gamma=\frac{1}{4}\pi$ modulo $\pi/2$.

\subsection{Inversion-symmetry breaking interface}
\label{activate_tunnel}

We next consider interfaces that break inversion symmetry rather than time-reversal symmetry. A potential barrier on the S side of the interface couples $\pm k_{\rm F}$, and thereby switches the $\tau_z$ index. This is modeled by a tunnel Hamiltonian of the form $H_{\rm interf}=V_{\rm barrier}\tau_\alpha$ with $\alpha\in\{x,y\}$, which preserves time-reversal symmetry ($H_{\rm interf}=\tau_x\sigma_y H_{\rm interf}^\ast\tau_x\sigma_y$).

The choice $H_{\rm interf}=V_{\rm barrier}\tau_x$ gives the transfer matrix
\begin{equation}
\Xi=e^{-\gamma'\tau_y},\;\;\gamma'=V_{\rm barrier}\ell/v_{\rm F}\lesssim V_{\rm barrier}/\Delta_0.\label{Xibetaprime}
\end{equation}
This preserves inversion symmetry [see Eq.\ \eqref{Xiinversion}], and does not activate Andreev reflection: $r_{he}=0$ for all $E$.

If instead we take the Hamiltonian $H_{\rm interf}=V_{\rm barrier}\tau_y$, we have $\Xi=e^{\gamma'\tau_x}$. Inversion symmetry is broken, and we find activated Andreev reflection:
\begin{align}
r_{he}=\frac{2i\sin\alpha\sinh 2\gamma'\cos\theta  }{\sin^2\alpha\sinh^2 2\gamma' \sin^2\theta+\left(\sin\alpha \cosh 2 \gamma'-i\cos\alpha \right)^2}.\label{reerheresult2}
\end{align}
At the Fermi level, and for $m_0\ll \beta\ll t_z$, the Andreev reflection probability is
\begin{equation}
R_{he}=\frac{4\sinh^2 2\gamma'}{\cosh^4 2\gamma'}.\label{Rheresult2}
\end{equation}
It reaches a maximum of unity for $\gamma'=\frac{1}{2}\ln(1+\sqrt 2)=0.441$, decaying to zero for both smaller and larger $\gamma'$.

\section{Conductance of the NS junction}
\label{sec_cond}

The reflection probabilities $R_{ee}=|r_{ee}|^2$ and $R_{he}=|r_{he}|^2$ determine the differential conductance $dI/dV=G(eV)$ of the NS junction, per unit surface area, according to\cite{BTK}
\begin{equation}
G(E)=\frac{e^2}{h}\int \frac{dk_y}{2\pi} \int \frac{dk_z}{2\pi} (1-R_{ee}+R_{he}).\label{GRhedef}
\end{equation}
The reflection amplitudes $r_{ee}$ and $r_{he}$, as a function of energy $E$ and transverse momentum components $k_y,k_z$, follow from the solution of Eq.\ \eqref{reflection_equations}, suitably generalized to include an arbitrary angle of incidence. 

We consider an incident electron near the Weyl point at $\bm{k}=(0,0,K)$, with $K\approx\beta/t_z\ll 1$. [The other Weyl cone at $-K$ gives the same contribution to the conductance and we may set $\theta=0$ in the transfer matrix \eqref{curlyTresult}.] We take $\mu_{\rm W},E>0$ so the electron is above the Fermi level at energy $\mu_{\rm W}+E$ in the upper half of the Weyl cone. The Andreev reflected hole is below the Fermi level at energy $\mu_{\rm W}-E$, which drops into the lower half of the Weyl cone when $E>\mu_{\rm W}$. For brevity we denote $q_x=t_x k_x$, $q_y=t_y k_y$, $q_z=t_z k_z-\beta$. 

We normalize the conductance by the total number $N(E)$ of propagating electron modes in the Weyl cones at energy $E$ above $\mu_{\rm W}$, given by
\begin{align}
N(E){}&=2\int \frac{dq_y}{2\pi t_y} \int \frac{dq_z}{2\pi t_z}\,\Theta\left[(E+\mu_{\rm W})^2-q_y^2-q_z^2\right]\nonumber\\
&{}=\frac{(E+\mu_{\rm W})^2}{2\pi t_y t_z}.\label{G0def}
\end{align}
(The prefactor 2 sums the contributions from the two Weyl cones.)

The low-energy Hamiltonian $H_K$ follows upon projection of the Hamiltonian \eqref{Htransformed} on the $\tau=-1$ band and expansion around the Weyl point,
\begin{equation}
H_{K}=-\nu_z(\sigma_x q_x+\sigma_y q_y)-\nu_0\sigma_z q_z-\mu_{\rm W}\nu_z\sigma_0.
\end{equation}
The $x$-component of the momentum is $- q_x$ and $+q_x$ for the incident and reflected electron, and $q'_x$ for the hole, with
\begin{equation}
\begin{split}
&q_x=\sqrt{(E+\mu_{\rm W})^2-q_y^2-q_z^2},\\
&q'_x={\rm sign}\,(E-\mu_{\rm W})\sqrt{(E-\mu_{\rm W})^2-q_y^2-q_z^2}.
\end{split}\label{kxkxprimedef}
\end{equation}
Only real $q_x$ contribute to the wave vector integration in Eq.\ \eqref{GRhedef}, and when $q'_x$ becomes imaginary one should set $R_{he}\equiv 0$. 
 
Substitution of the corresponding spinors into Eq.\ \eqref{reflection_equations} (normalized to unit flux) gives the mode matching condition
\begin{align}
&\sqrt{\frac{q_x(E+\mu_{\rm W}-q_z)}{q'_x(E-\mu_{\rm W}-q_z)}}\begin{pmatrix}
b\\
-ib\\
(q'_x+iq_y)r_{he}\\
(E-\mu_{\rm W}-q_z)r_{he}
\end{pmatrix}=\nonumber\\
&\qquad=
{\cal T}
\begin{pmatrix}
a\\
ia\\
q_x-iq_y+(q_x+iq_y)r_{ee}\\
(E+\mu_{\rm W}-q_z)(1-r_{ee})
\end{pmatrix}.\label{reflection_equations2}
\end{align}

\begin{figure}[tb]
\centerline{\includegraphics[width=0.8\linewidth]{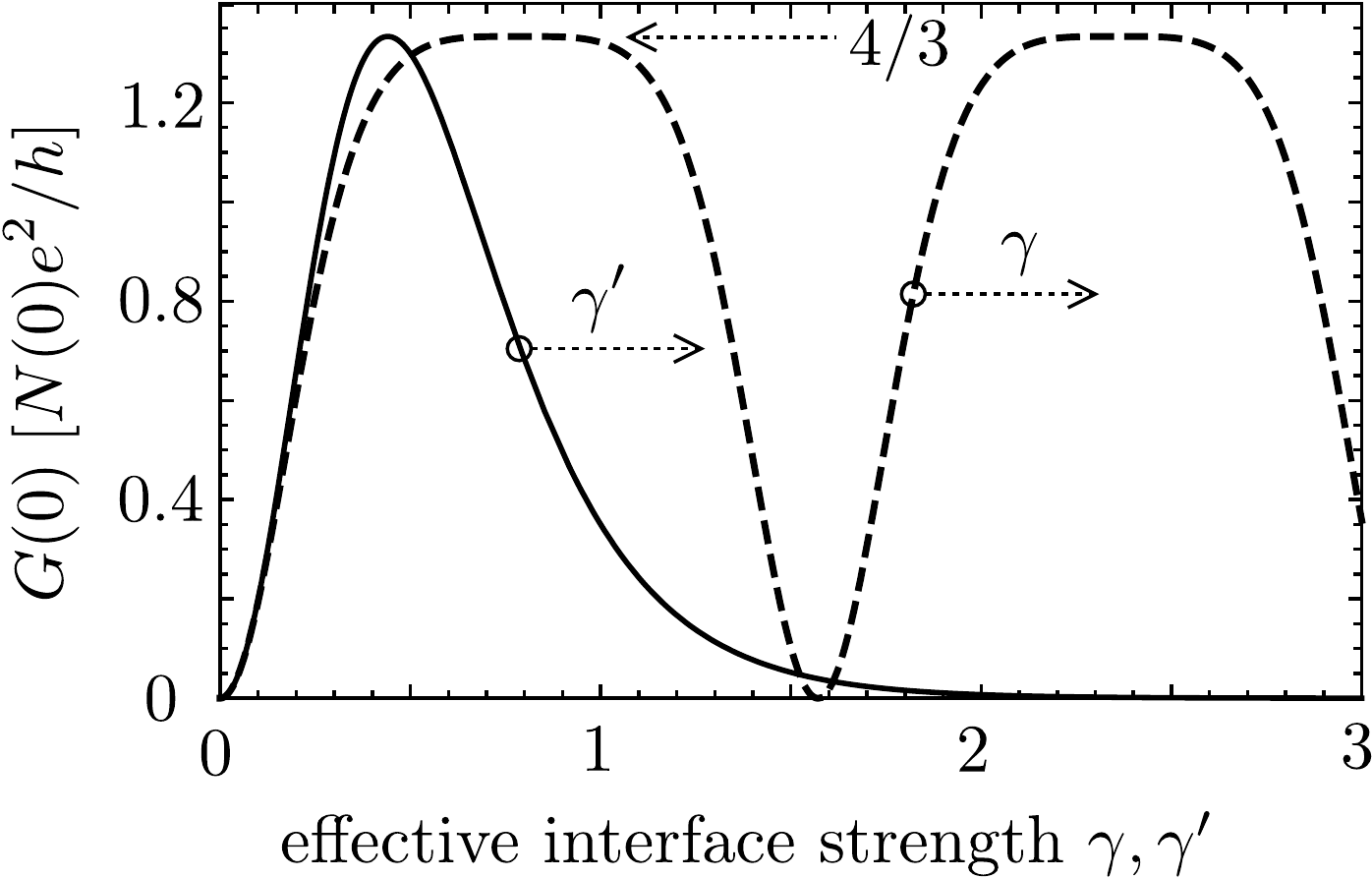}}
\caption{Zero-bias conductance of the NS junction, calculated from Eq.\ \eqref{Gzerobias}, for the spin-active interface (dashed curve) and for the inversion-symmetry breaking interface (solid curve). The conductance is normalized by the number of modes $N$ from Eq.\ \eqref{G0def}. For the inactive interface the conductance vanishes.
}
\label{fig_zerobiasG}
\end{figure}

For the inactive interface, when $\Xi=1$, the Andreev reflection amplitude vanishes at all energies for all angles of incidence. Andreev reflection is activated by the spin-active interface or by the inversion-symmetry-breaking interface, as discussed in Sec.\ \ref{ARactivation}. At the Fermi level ($E=0$, $q'_x=-q_x$) we recover the results \eqref{Rheresult} and \eqref{Rheresult2} multiplied by the factor $q_x^2/(q_x^2+q_z^2)$ that accounts for the deviation from normal incidence. The resulting zero-bias conductance is given by
\begin{equation}
\lim_{V\rightarrow 0}\frac{dI}{dV}=\tfrac{16}{3}N(0)\frac{e^2}{h}\times\begin{cases}
\sin^2 2\gamma/(1+\sin^2 2\gamma)^{2},\\
\sinh^2 2\gamma'/\cosh^{4} 2\gamma',
\end{cases}\label{Gzerobias}
\end{equation}
as plotted in Fig.\ \ref{fig_zerobiasG}, with $\gamma=E_{\rm Zeeman}\ell/v_{\rm F}\lesssim E_{\rm Zeeman}/\Delta_0$ in the spin-active interface Hamiltonian $H_{\rm interf}=E_{\rm Zeeman}\sigma_y$, and $\gamma'=V_{\rm barrier}\ell/v_{\rm F}\lesssim V_{\rm barrier}/\Delta_0$ in the inversion-symmetry breaking case $H_{\rm interf}=V_{\rm barrier}\tau_y$.

\begin{figure}[tb]
\centerline{\includegraphics[width=0.8\linewidth]{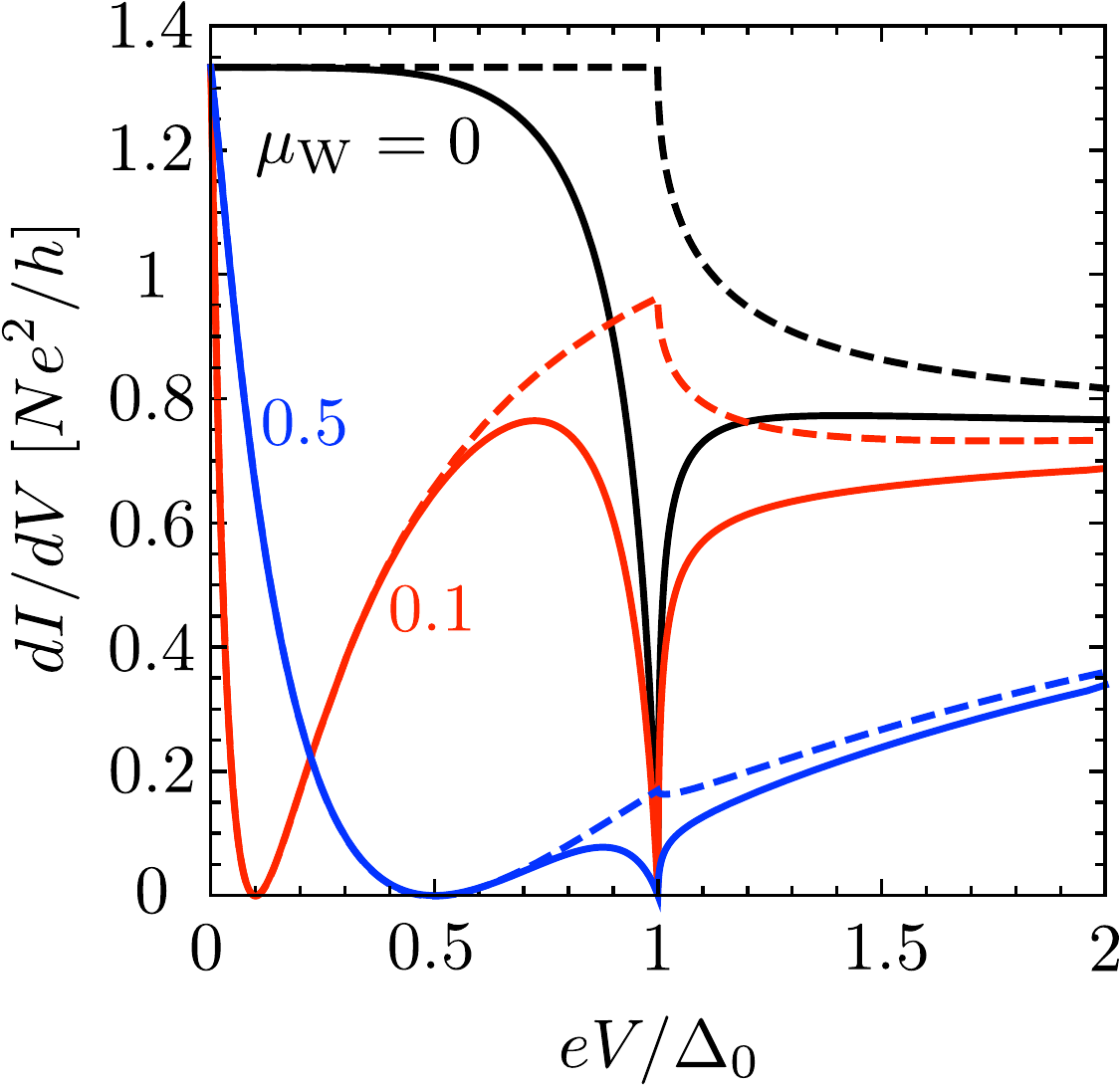}}
\caption{Differential conductance of the NS junction, calculated from Eqs.\ \eqref{GRhedef} and \eqref{reflection_equations2}, for the spin-active interface of Sec.\ \ref{activate_spin} (dashed curves, for $H_{\rm interf}=E_{\rm Zeeman}\sigma_y$ with $\gamma=\pi/4$), and for the inversion-symmetry breaking interface of Sec.\ \ref{activate_tunnel} (solid curves, for $H_{\rm interf}=V_{\rm barrier}\tau_y$ with $\gamma'=\frac{1}{2}\ln(1+\sqrt{2})$). For $eV\gg\Delta_0$, all curves tend to the normal-state interface conductance of $0.8\,Ne^2/h$.
}
\label{fig_conductance}
\end{figure}

The voltage-dependent differential conductance is plotted in Fig.\ \ref{fig_conductance}. The conductance vanishes at $eV=\mu_{\rm W}<\Delta_0$, when the hole touches the Weyl point. (The same feature appears at the Dirac point in graphene \cite{Bee06}.)

\section{Weyl semimetal -- Weyl superconductor junction}
\label{WeylWeylsec}

So far we have considered the junction between a Weyl semimetal and a superconductor formed from a conventional metal. A doped Weyl semimetal can itself become superconducting, forming a Weyl superconductor.\cite{Has17,Yan17} In this section we study how the chirality blockade manifests itself in an NS junction between the normal and superconducting state of Weyl fermions. To make contact with a specific microscopic model, we consider the heterostructure approach of Burkov and Balents \cite{Bur11}, which can describe both a Weyl semimetal and a Weyl superconductor.\cite{Men12,Bed15}

\subsection{Heterostructure model}

For the Weyl semimetal, we start from a multilayer heterostructure, composed of layers of a magnetically doped topological insulator (such as ${\rm Bi}_2{\rm Se}_3$), separated by a normal-insulator spacer layer with periodicity $d$. Its Hamiltonian is\cite{Bur11,Zha09,Fu10}
\begin{align}
H(\bm{k})={}&v_{\rm F}\tau_z(-\sigma_y   k_x +\sigma_x   k_y )+\beta\tau_0\sigma_z\nonumber\\
&+(m_k\tau_x  -\tau_y t_z  \sin k_z d)\sigma_0,\label{HWBB}\\
m_k={}&t'_z+t_z\cos k_z d.\nonumber
\end{align}
The Pauli matrices $\sigma_i$ act on the spin degree of freedom of the surface electrons in the topological insulator layers. The $\tau_z=\pm 1$ index distinguishes the orbitals on the top and bottom surfaces, coupled by the $t'_z$ hopping within the same layer and by the $t_z$ hopping from one layer to the next. Magnetic impurities in the topological insulator layers produce a perpendicular magnetization, leading to an exchange splitting $\beta$. The two Weyl points are at $\bm{k}=(0,0,\pi/d\pm K)$, with
\begin{equation}
K^2\approx\frac{{\beta^2-(t_z-t'_z)^2}}{{d^2 t_z t'_z}}.\label{KapproxBB}
\end{equation}
They are closely spaced near the edge of the Brillouin zone for $|t_z-t'_z|\ll\beta\ll t_z d$.

To make contact with the generic Weyl Hamiltonian \eqref{HW}, we note the unitary transformation
\begin{align}
U_0 H(\bm{k})U_0^\dagger={}&v_{\rm F}\tau_z(\sigma_x k_x +\sigma_y k_y) - \tau_z \sigma_z t_z \sin k_z d\nonumber\\
&+m_k\tau_x\sigma_0+\beta\tau_0\sigma_z,\label{HMBtransformed}\\
U_0={}&\exp[-\tfrac{1}{4}i\pi(\tau_0+\tau_x)\sigma_z].\nonumber
\end{align}
We will make use of this transformation later on.

\begin{figure}[tb]
\centerline{\includegraphics[width=0.7\linewidth]{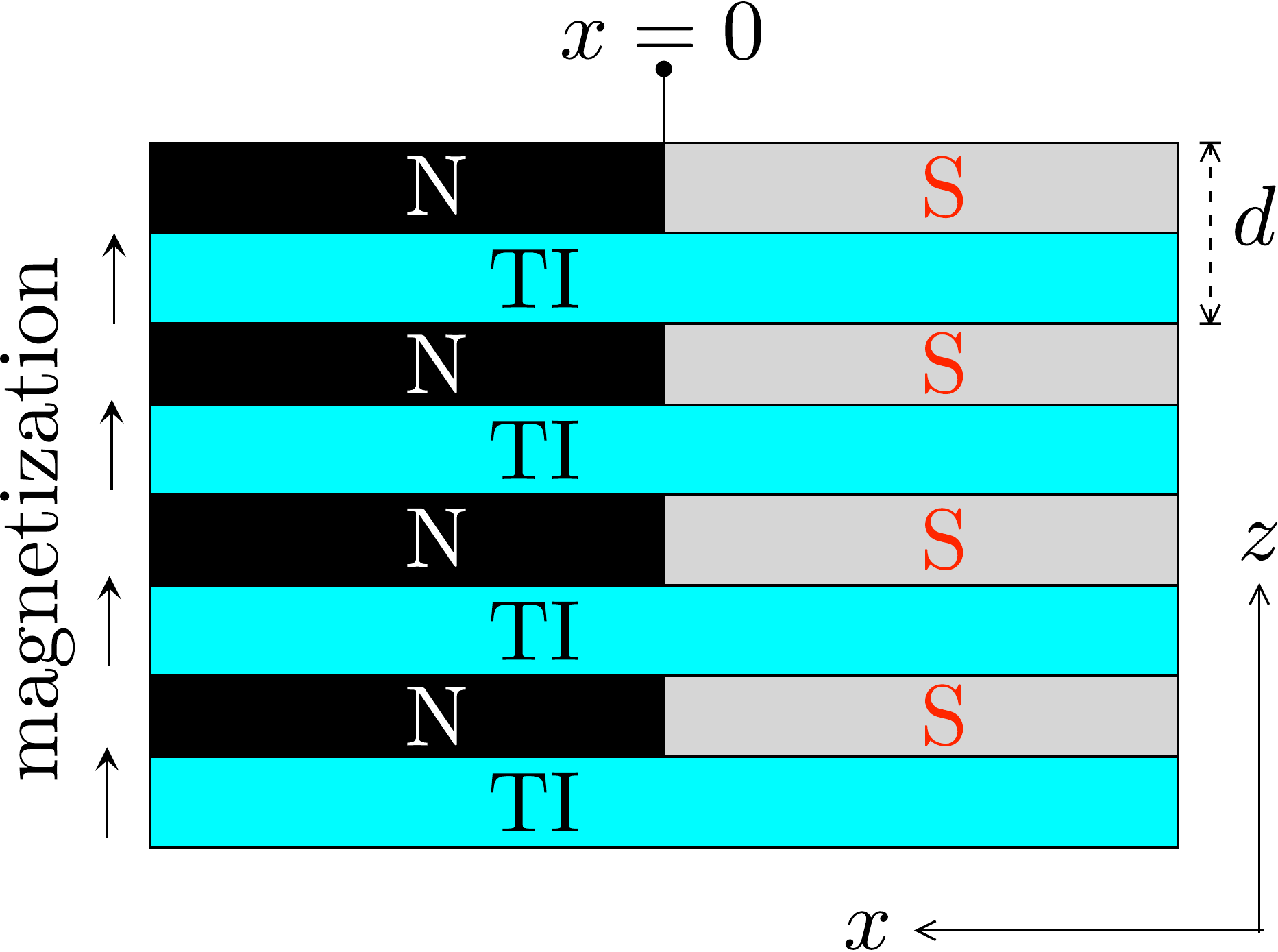}}
\caption{Cross-section through a layered Weyl semimetal-superconductor junction, based on the heterostructure model\cite{Bur11,Men12} of alternating topological insulator (TI) layers and normal (N) or superconducting (S) spacer layers. In this model the orbital $\tau$ degree of freedom refers to the conducting top and bottom surfaces of the TI layers.
}
\label{fig_heterostructure}
\end{figure}

Following Meng and Balents \cite{Men12}, the spacer layer may have a spin-singlet \textit{s}-wave pair potential $\Delta$, with a uniform phase throughout the heterostructure (which we set to zero, allowing us to take $\Delta$ real). The pair potential induces superconductivity in the top and bottom surfaces of the topological insulator layers, as described by the BdG Hamiltonian
\begin{align}
{\cal H}(\bm{k})={}&v_{\rm F}\nu_z\tau_z(-\sigma_y   k_x +\sigma_x   k_y )+\beta\nu_0\tau_0\sigma_z\nonumber\\
&+\nu_z(m_k\tau_x  -\tau_y  t_z\sin k_z d )\sigma_0-\mu\nu_z\tau_0\sigma_0+\bm{\Delta},\nonumber\\
\bm{\Delta}={}&\Delta(x)\nu_x\tau_0\sigma_0.\label{HSMB}
\end{align}
It acts on eight-component Nambu spinors $\Psi$ with elements 
\begin{equation}
\Psi=(\psi_{+\uparrow},\psi_{+\downarrow},\psi_{-\uparrow},\psi_{-\downarrow},\psi^\ast_{+\downarrow},-\psi^\ast_{+\uparrow},\psi^\ast_{-\downarrow},-\psi^\ast_{-\uparrow}),\label{Psidef}
\end{equation}
where $\pm$ refers to the top and bottom surface and $\updownarrow$ refers to the spin band.

The pair potential $\bm{\Delta}$ in Eq.\ \eqref{HSMB} is diagonal in the $\tau$ and $\sigma$ degrees of freedom. The corresponding BCS pairing interaction,
\begin{align}
H_{\rm BCS}={}&\Delta\sum_{\bm{k}}\left[c^{\dagger}_{+\uparrow}(\bm{k})c^{\dagger}_{+\downarrow}(-\bm{k})+c^{\dagger}_{-\uparrow}(\bm{k})c^{\dagger}_{-\downarrow}(-\bm{k})\right]\nonumber\\
&+{\rm H.c.,}\label{HintBCS}
\end{align}
represents zero-momentum pairing of spin-up and spin-down electrons within the same conducting surface of each topological insulator layer (inversion-symmetric, spin-singlet, intra-orbital pairing).

The BCS pairing interaction \eqref{HintBCS} corresponds to a scalar pair potential in the spin and orbital degrees of freedom. We restrict ourselves to that pairing symmetry in this section. Other BCS pair potentials (spin-triplet and pseudoscalar spin-singlet) are considered in Appendix\ \ref{otherpairing}.

To describe a NS interface at $x=0$, we set $\Delta(x)=0$ for $x>0$ and $\Delta(x)=\Delta_0$ for $x<0$ (see Fig.\ \ref{fig_heterostructure}). We also adjust the chemical potential $\mu(x)$, from a small value $\mu_{\rm W}$ for $x>0$ to a large value $\mu_{\rm S}$ for $x<0$. For the other parameters we take $x$-independent values.
 
\subsection{Mode matching at the NS interface}

We can now follow the mode-matching analysis of the preceding sections, with one simplification and one complication. The simplification is that, because we have the same Weyl Hamiltonian on the two sides of the NS interface, we no longer need an interface matrix to conserve current across the interface. The complication is that the block-diagonalization in the $\tau$ degree of freedom on the N side of the interface introduces off-diagonal blocks in the pair potential on the S side.

The unitary transformation that achieves this partial block-diagonalization is
\begin{align}
&\tilde{\cal H}={\cal V}{\cal H}{\cal V}^\dagger,\;\;{\cal V}=\begin{pmatrix}
\tau_y\sigma_z\Omega_{\theta}U_{0}&0\\
0&\Omega_{\theta}U_{0}
\end{pmatrix},\label{UBBdef}\\
&\Omega_\theta=\exp(-\tfrac{1}{2}i\theta\tau_y\sigma_z),\nonumber
\end{align}
with $U_0$ from Eq.\ \eqref{HMBtransformed}. The $k_z$-dependent angle $\theta$ is defined by
\begin{equation}
\begin{split}
&\cos\theta=(t_z\sin k_z d)/M_{k},\;\;\sin\theta=m_{k}/M_{k},\\
&M_{k}=\sqrt{m_k^2+t_z^2\sin^2 k_z d}.
\end{split}\label{MkdefBB}
\end{equation}
For closely-spaced Weyl points (when $|t_z-t'_z|\ll\beta\ll t_z d$) we may approximate $\sin\theta\approx 0$, $|\cos\theta|\approx  1$. 

The transformed Hamiltonian is
\begin{align}
\tilde{\cal H}(\bm{k})={}&v_{\rm F}\nu_z\tau_z(\sigma_x  k_x+\sigma_y  k_y)+M_{k}\nu_0\tau_z\sigma_z+\beta\nu_0\tau_0\sigma_z\nonumber\\
&-\mu\nu_z\tau_0\sigma_0+\tilde{\bm{\Delta}},\nonumber\\
\tilde{\bm{\Delta}}\equiv{}&{\cal V}\bm{\Delta}{\cal V}^\dagger=\Delta(x)\nu_x\tau_y\sigma_z.\label{HtransformedMB}
\end{align}
This has the same block-diagonal form \eqref{Htransformed} on the N side $x>0$ of the interface (where $\Delta=0$), but on the S side $x<0$ the transformed pair potential $\tilde{\bm{\Delta}}$ is off-diagonal in the $\tau$ degree of freedom.\cite{MBnote}

We again assume $\mu_{\rm S}\gg\mu_{\rm W}$ so that in S we may neglect the transverse wave vector component $k_y$ and take $k_z$ at the Weyl point, where $M_k=\beta$. The wave equation in S corresponding to the Hamiltonian \eqref{HtransformedMB} then reads
 \begin{align}
&v_{\rm F}\frac{\partial}{\partial x}\psi(x)=X_{\rm S}\psi(x),\;\;x<0,\label{XSdef}\\
&X_{\rm S}=i(E\nu_z+\mu_{\rm S}\nu_0)\tau_z\sigma_x-\beta\nu_z(\tau_0+\tau_z)\sigma_y-\Delta_0\nu_y\tau_x\sigma_y.\nonumber
\end{align}
As derived in Appendix\ \ref{app_boundarycond}, the decaying eigenvectors for $E<\Delta_0$ and $x\rightarrow -\infty$ satisfy 
\begin{equation}
\nu_x\tau_y\sigma_z\psi= \exp(i\alpha \nu_z\tau_z\sigma_x)\psi,\label{psiMBcondition}
\end{equation}
with $\alpha={\rm arccos}(E/\Delta_0)\in(0,\pi/2)$. The corresponding boundary condition on $\psi=(\psi_e,\psi_h)$ is
\begin{equation}
\psi_h(0)={\cal T}\psi_e(0),\;\;{\cal T}= e^{i\alpha\tau_z\sigma_x}\tau_y\sigma_z.\label{TMB}
\end{equation}
Because $\psi(x)$ is now continuous across the interface, we do not need to distinguish $0^+$ and $0^-$ as we needed to do in Sec.\ \ref{sec_boundarycond}.

Substitution of ${\cal T}$ into the mode matching equation \eqref{reflection_equations2} gives $r_{he}\equiv 0$; fully suppressed Andreev reflection at all energies and all angles of incidence. This is the chirality blockade.

\section{Fermi-arc mediated Josephson effect}
\label{FermiJosephson}

While the conductance of a single NS interface is fully suppressed by the chirality blockade, the supercurrent through an SNS junction is nonzero because of overlapping surface states (Fermi arcs) on the two NS interfaces. We have calculated this Fermi-arc mediated Josephson effect (see Appendix\ \ref{FermiarcJcalc}), and we summarize the results.

The Fermi arcs connect the Weyl cones of opposite chirality \cite{Wan11}. As they pass through the center of the Brillouin zone, the chirality blockade is no longer operative and the Fermi arcs acquire a mixed electron-hole character. At $k_z=0$, the surface states are charge neutral Majorana fermions.\cite{Bai17}

\begin{figure}
\includegraphics[width=0.9\linewidth]{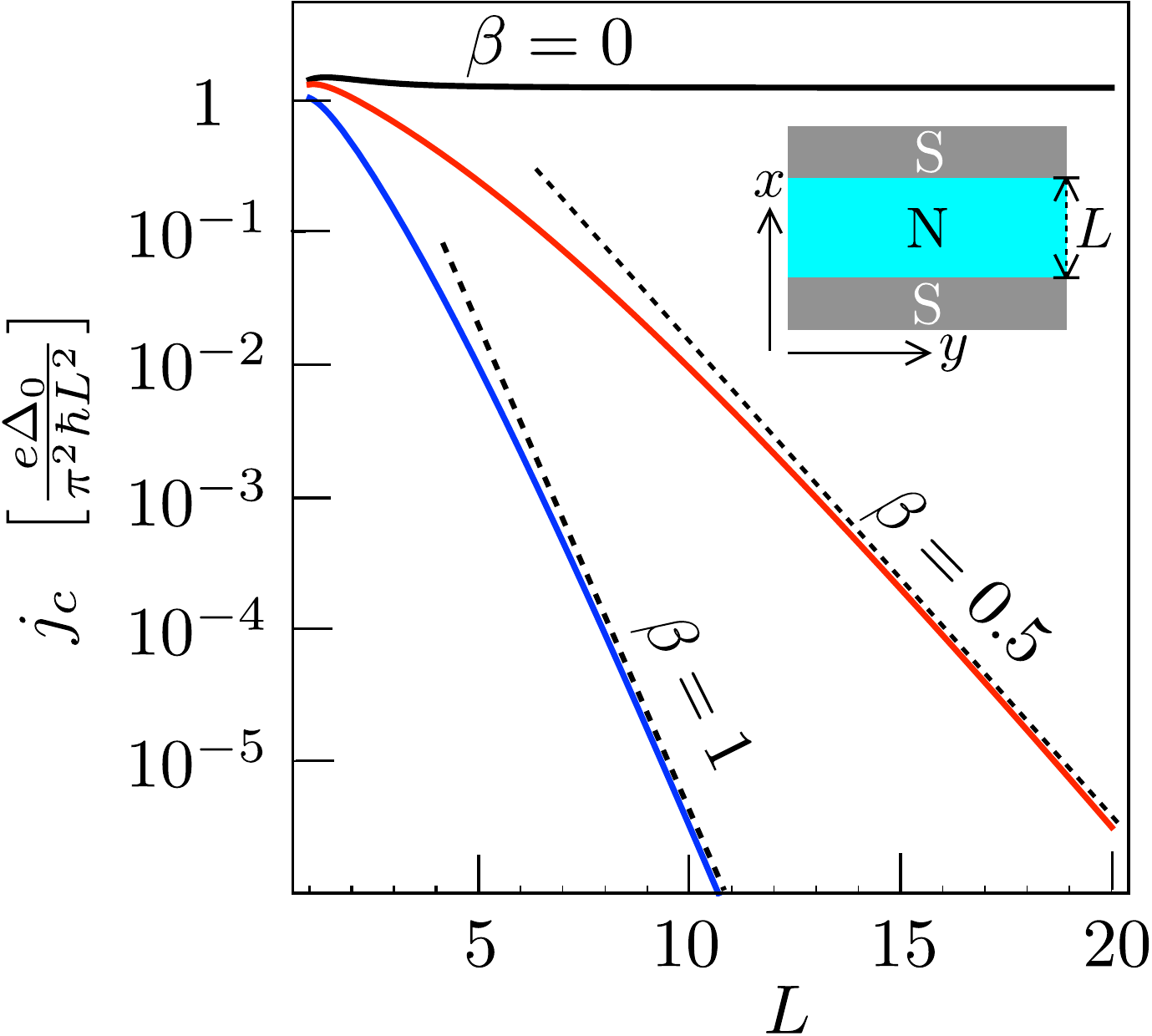}
\caption{Critical current density $j_c$ of the SNS junction as a function of the separation $L$ of the NS interfaces  for different values of $\beta$, calculated from the Hamiltonian \eqref{HSMB} for $\mu=0$, $v_{\rm F}=t_z=t_z'=d=1$. The dashed lines indicate the exponential decay $\propto e^{-c\beta L/v_{\rm F}}$ with $c=1.7$.
} 
\label{Jcurrent}
\end{figure}

The Fermi arcs are bound to the NS interface over a distance of order $v_{\rm F}/\beta$, so a coupling of the two NS interfaces is possible if their separation $L\lesssim v_{\rm F}/\beta$. For larger $L$, the critical current is suppressed $\propto \exp(-L/\xi_{\rm arc})$, with $\xi_{\rm arc}\simeq v_{\rm F}/\beta$ the penetration depth of the surface Fermi arc into the bulk (see Fig.\ \ref{Jcurrent}).

\section{Discussion}
\label{sec_discuss}

In conclusion, we have shown that Andreev reflection at the interface between a Weyl semimetal and a spin-singlet \textit{s}-wave superconductor is suppressed by a mismatch of the chirality of the incident electron and the reflected hole. Zero-momentum (\textit{s}-wave) pairing requires that the electron and hole have opposite chirality, while singlet pairing requires that they occupy opposite spin bands, and these two requirements are incompatible, as illustrated in Fig.\ \ref{fig_layout}.

We have identified two mechanisms that can remove the chirality blockade and activate Andreev reflection. The first mechanism, a spin-active interface, has the same effect as spin-triplet pairing: it enables Andreev reflection by allowing an electron and a hole to be in the same spin band. The second mechanism, inversion-symmetry breaking either at the interface or in the pair potential, is more subtle, as we now discuss.

\begin{figure}[tb]
\centerline{\includegraphics[width=0.8\linewidth]{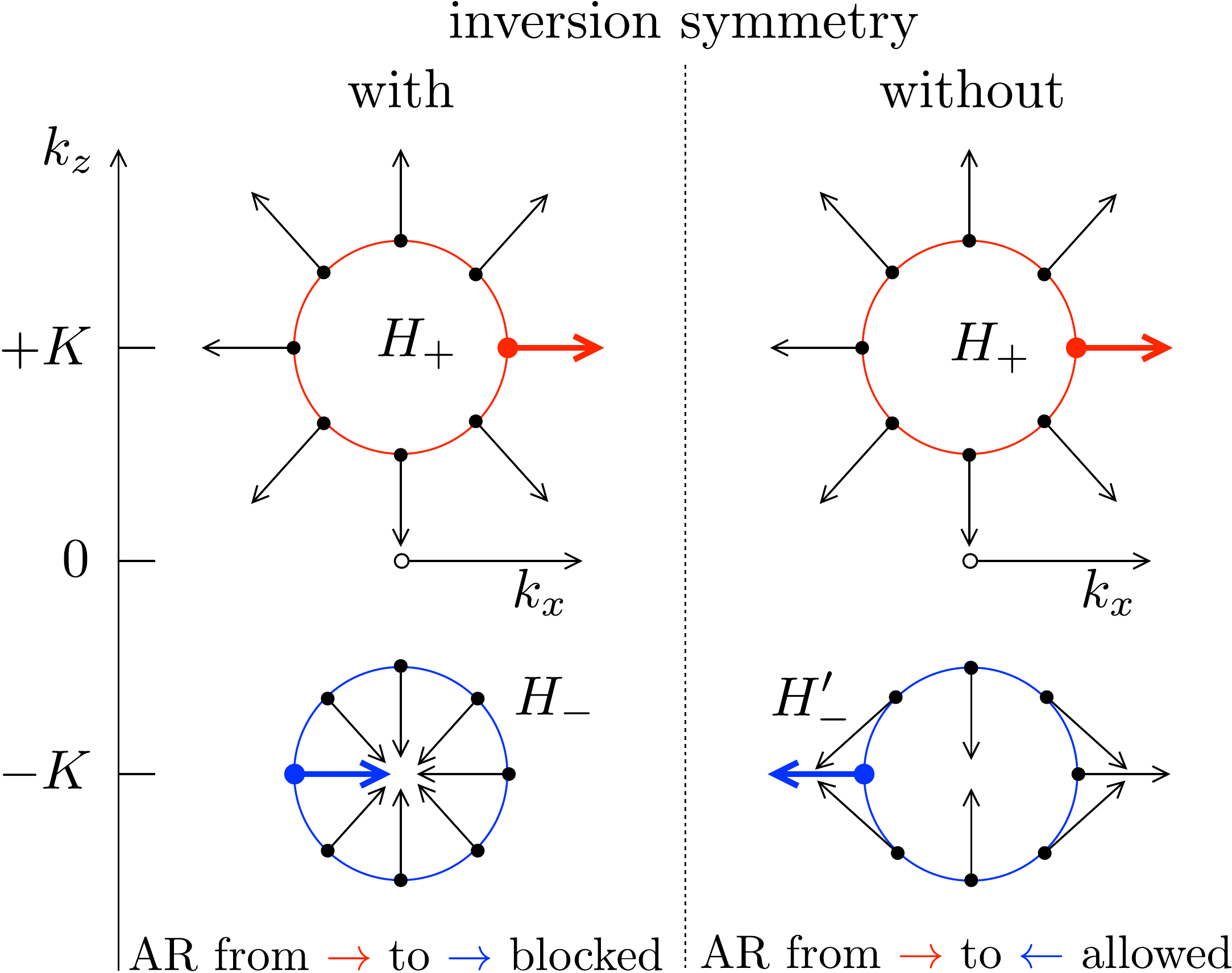}}
\caption{Illustration of the spin-momentum locking for states at the Fermi energy in a pair of Weyl cones at $\bm{k}=(0,0,\pm K)$. The arrows indicate the direction of the spin polarization for a momentum eigenstate at $k_y=0$, as a function of $k_x$ and $k_z$. The left column is for the Hamiltonian $H_{\rm W}=(H_+,H_-)$ with inversion symmetry, the right column is for $H'_{\rm W}=(H_+,H'_-)$ without inversion symmetry. Andreev reflection (AR) along the $x$-direction on a superconductor with zero-momentum spin-singlet pairing is blocked for $H_{\rm W}$ (the red and blue arrows point in the same direction, so the spin is not inverted, as it should be for spin-singlet pairing), while it is allowed for $H'_{\rm W}$ (red and blue arrows point in opposite directions).
}
\label{fig_spintexture}
\end{figure}

Consider the single-cone Weyl Hamiltonian centered at $\bm{k}=(0,0,+K)$,
\begin{equation}
H_{+}=v_x k_x\sigma_x+v_y k_y\sigma_y+v_z (k_z-K)\sigma_z.\label{Hplusdef}
\end{equation}
By definition, its chirality is $C={\rm sign}\,(v_x v_y v_z)$. For the second Weyl cone centered at $\bm{k}=(0,0,-K)$ of opposite chirality, we can take either
\begin{equation}
H_{-}=-v_x k_x\sigma_x-v_y k_y\sigma_y-v_z (k_z+K)\sigma_z\label{Hmindef}
\end{equation}
or 
\begin{equation}
H'_{-}=v_x k_x\sigma_x+v_y k_y\sigma_y-v_z (k_z+K)\sigma_z,\label{Hminprimedef}
\end{equation}
or some permutation of $x,y,z$, but either all three signs or one single sign of the velocity components must flip. The first choice satisfies inversion symmetry, $H_-(-\bm{k})=H_+(\bm{k})$, while the second choice does not. In Fig.\ \ref{fig_spintexture} we show the spin-momentum locking in the pair of Weyl cones $H_{\rm W}=(H_+,H_-)$ and $H'_{\rm W}=(H_+,H'_-)$ with and without inversion symmetry. We see that the chirality blockade can be removed by breaking inversion symmetry.

This explains why Uchida, Habe, and Asano \cite{Uch14} (who, with Cho, Bardarson, Lu, and Moore \cite{Cho12}, fully appreciated the importance of spin-momentum locking for superconductivity in a Weyl semimetal) did not find any suppression of Andreev reflection at normal incidence on the NS interface. Their two-band model of a Weyl semimetal\cite{Yan11,Oga14,twoband} has the same spin texture as $H'_{\rm W}$ --- hence it breaks inversion symmetry and does not show the chirality blockade. The relevance of inversion symmetry also explains why no chirality blockade appeared in Refs.\ \onlinecite{Che13,Kha16,Mad17}, where a pseudoscalar pair potential was used that breaks this symmetry (see Appendix\ \ref{app_inversionbroken}).

The chirality blockade suppresses the superconducting proximity effect, but since it can be lifted in a controlled way by a Zeeman field (see Fig.\ \ref{fig_zerobiasG}), it offers opportunities for spintronics applications. In the geometry of Fig.\ \ref{fig_layout}, a magnetic field in the $y$-direction, in the plane of the NS interface and perpendicular to the magnetization, activates Andreev reflection when the Zeeman energy $E_{\rm Zeeman}$ becomes comparable to the superconducting gap $\Delta_0$. (To prevent pair-breaking effects from this Zeeman field one can use a thin-film superconductor with strong spin-orbit coupling \cite{Nam16}.) For a typical Zeeman energy of 1~meV/Tesla and a typical gap of 0.1~meV, a 100~mT magnetic field can then activate the transfer of Cooper pairs through the NS interface. This provides a phase-insensitive alternative to the phase-sensitive control of Cooper pair transfer in a Josephson junction.

\acknowledgments
We have benefited from discussions with Tobias Meng.
This research was supported by the Netherlands Organization for Scientific Research (NWO/OCW) and an ERC Synergy Grant.

\appendix

\section{Derivation of the boundary condition at a Weyl semimetal -- Weyl superconductor interface}
\label{app_boundarycond}

Equation\ \eqref{psiSboundary} gives the effective boundary condition at the NS interface between a Weyl semimetal and a conventional superconductor. Here we generalize this to the interface between a Weyl semimetal and a Weyl superconductor. We allow for a more general pairing symmetry than considered in the main text, and in Appendix\ \ref{otherpairing} we apply the boundary condition to spin-triplet pairings and to a pseudoscalar spin-singlet pairing.

As discussed in the related context of graphene \cite{Tit06}, the \textit{local} coupling of electrons and holes at the NS interface that is expressed by the effective boundary condition holds under three conditions: (i) The chemical potential $\mu_{\rm S}$ in the superconducting region is the largest energy scale in the problem, much larger than the superconducting gap $\Delta_0$ and much larger than the chemical potential $\mu_{\rm N}$ in the normal region; (ii) the interface is smooth and impurity-free on the scale of the superconducting coherence length $\hbar v_{\rm F}/\Delta_0$; and (iii) there is no lattice mismatch at the NS interface.

We start from Eq.\ \eqref{XSdef}, which governs the decay of the wave function in the superconducting region,
 \begin{align}
&v_{\rm F}\frac{\partial}{\partial x}\psi(x)=X_{\rm S}\psi(x),\;\;x<0,\label{XSdefapp}\\
&X_{\rm S}=(i\mu_{\rm S}\tau_z\sigma_x+Y_{\rm S}),\;\;Y_{\rm S}=i\nu_z\tau_z\sigma_x(E-\tilde{\bm{\Delta}}).\nonumber
\end{align}
We have omitted the $\beta$ term, which anticommutes with the $\mu_{\rm S}$ term and can be neglected in the large-$\mu_{\rm S}$ limit. We seek a boundary condition on $\psi$ at $x=0$ that ensures decay for $x\rightarrow -\infty$.

In the most general case, $\tilde{\bm{\Delta}}$ is an Hermitian $8\times 8$ matrix that satisfies the electron-hole symmetry relation
\begin{equation}
\nu_y\sigma_y\tilde{\bm{\Delta}}^\ast\nu_y\sigma_y=-\tilde{\bm{\Delta}}.\label{Deltasymmapp}
\end{equation}
 We make the following four additional assumptions:
\begin{enumerate}
\item $\tilde{\bm{\Delta}}$ anticommutes with $\nu_z$ (so it is fully off-diagonal in the electron-hole degree of freedom);
\item $\tilde{\bm{\Delta}}$ commutes with $\tau_z\sigma_x$ (anticommuting terms do not contribute to the spectrum of $X_{\rm S}$ in the large-$\mu_{\rm S}$ limit, so they may be ignored);
\item $\tilde{\bm{\Delta}}$ is independent of the momentum perpendicular to the NS interface (it may depend on the parallel momentum);
\item $\tilde{\bm{\Delta}}$ squares to a scalar $\Delta_0^2$ (this assumption is not essential, but allows for a simple closed-form answer).
\end{enumerate}
Under these conditions $X_{\rm S}$ and $Y_{\rm S}$ commute, so they can be diagonalized simultaneously. Moroever, $Y_{\rm S}^2=\Delta_0^2-E^2$, hence a decaying wave function for $E<\Delta_0$ is an eigenfunction of $Y_{\rm S}$ with eigenvalue $+\sqrt{\Delta_0^2-E^2}$,\begin{equation}
Y_{\rm S}\psi=\sqrt{\Delta_0^2-E^2}\,\psi.\label{YSrelation}
\end{equation}

We rearrange this to obtain a relation between the electron and hole components of $\psi=(\psi_e,\psi_h)$:
\begin{align}
&-i\nu_z\tau_z\sigma_x\tilde{\bm{\Delta}}\psi=\left(-iE\nu_z\tau_z\sigma_x+\sqrt{\Delta_0^2-E^2}\right)\psi\nonumber\\
&\Rightarrow\tilde{\bm{\Delta}}\psi=\left(E+i\sqrt{\Delta_0^2-E^2}\,\nu_z\tau_z\sigma_x\right)\psi\nonumber\\
&\Rightarrow \tilde{\bm{\Delta}}\psi=\Delta_0\exp(i\alpha\nu_z\tau_z\sigma_x)\psi,\label{rearranged_relation}
\end{align}
with $\alpha={\rm arccos}\,(E/\Delta_0)\in(0,\pi/2)$. For a superconducting phase $\varphi$ we can decompose
\begin{equation}
\tilde{\bm{\Delta}}=\Delta_0(\nu_x\cos\varphi-\nu_y\sin\varphi)\bm{\chi},\label{deltachidecompose}
\end{equation}
with $\bm{\chi}$ a $4\times 4$ Hermitian matrix that squares to unity and commutes with $\tau_z\sigma_x$. We thus arrive at the desired boundary condition,
\begin{equation}
e^{i\varphi}\bm{\chi}\psi_h(0)=e^{i\alpha\tau_z\sigma_x}\psi_e(0).\label{psihpsiechi}
\end{equation}
In a more general geometry, with a unit vector $\bm{n}$ in the $x$--$y$ plane perpendicular to the NS interface and pointing from N to S, we can write the boundary condition as
\begin{equation}
\psi_h(0)={\cal T}\psi_e(0),\;\;{\cal T}=e^{-i\varphi}\exp\bigl[-i\alpha\tau_z(\bm{n}\cdot\bm{\sigma})\bigr]\bm{\chi}.
\label{psihpsiechi2}
\end{equation}

This was derived for subgap energies $E<\Delta_0$. The boundary condition still holds by analytic continuation for $E>\Delta_0$, when $\alpha=-i\,{\rm arcosh}\,(E/\Delta_0)$ is imaginary, provided that there is no particle current incident on the NS interface from the superconducting side.

\section{Generalizations to other pairing symmetries}
\label{otherpairing}

The pair potential $\bm{\Delta}=\Delta_0\nu_x\tau_0\sigma_0$ in the Meng-Balents Hamiltonian \eqref{HSMB} represents inversion-symmetric, spin-singlet, intra-orbital pairing, appropriate for the heterostructure model of Fig.\ \ref{fig_heterostructure}. Other types of pairing may be relevant for Weyl semimetals with intrinsic superconductivity.\cite{Bed15,Fu10} We calculate the corresponding Andreev reflection probabilities.

\subsection{Spin-triplet pair potential}
\label{app_spintriplet}

For the three $s=x,y,z$ spin-triplet pairings, the relationship between the pair potential $\bm{\Delta}_s$ in the Hamiltonian \eqref{HSMB} and the transformed pair potential $\tilde{\bm{\Delta}}_s$ in the Hamiltonian \eqref{HtransformedMB} is
\begin{subequations}\label{Deltatriplet}
\begin{align}
&\bm{\Delta}_s=\Delta_0\nu_x\tau_y\sigma_s\Rightarrow\tilde{\bm{\Delta}}_{s}=-\Delta_0\nu_y\bm{\chi}_s,\\
&\bm{\chi}_x=-\tau_0\sigma_x\cos \theta-\tau_y\sigma_y\sin \theta,\\
&\bm{\chi}_y=-\tau_0\sigma_y\cos \theta+\tau_y\sigma_x\sin \theta,\\
&\bm{\chi}_z=\tau_x\sigma_z\cos \theta-\tau_z\sigma_0\sin \theta.
\end{align}
\end{subequations}

Each $\bm{\chi}_s$ squares to unity but only $\bm{\chi}_x$ and $\bm{\chi}_z$ commute with $\tau_z\sigma_x$. The $s=y$ pairing anticommutes and does not open a gap in the large-$\mu_{\rm S}$ limit. For the $s=x$ and $s=z$ pairings we can read off the electron-hole coupling matrix ${\cal T}_s$ from Eq.\ \eqref{psihpsiechi2},
\begin{equation}
{\cal T}_s=-ie^{i\alpha\tau_z\sigma_x}\bm{\chi}_s,\label{chitriplet}
\end{equation}
 and then derive the Andreev reflection amplitude by solving Eq.\ \eqref{reflection_equations2}. The result for normal incidence is
 \begin{subequations}
\label{rhesxz}
\begin{align}
&r_{he}=\frac{2\sin\alpha\cos\theta}{\cos^2\theta-e^{2i\alpha}},\;\;\text{for}\;\;s=x,\\
&r_{he}=-\frac{2i\cos\alpha\sin\theta}{\sin^2\theta+e^{2i\alpha}},\;\;\text{for}\;\;s=z.
\end{align}
\end{subequations}

More generally, for any angle of incidence, we have at the Fermi level (when $E=0\Rightarrow\alpha=\pi/2$) the Andreev reflection probabilities
\begin{equation}
\begin{split}
&R_{he}=\frac{v_{\rm F}^2k_x^2}{\mu^2-v_{\rm F}^2 k_y^2}\frac{4\cos^2 \theta}{ (1+\cos^2\theta)^2},\;\;\text{for}\;\;s=x,\\
&R_{he}=0,\;\;\text{for}\;\;s=z.
\end{split}
\label{RheFermilevelspinxz}
\end{equation}

\subsection{Pseudoscalar spin-singlet pair potential}
\label{app_inversionbroken}

The pairing interaction
\begin{align}
H'_{\rm BCS}={}&\Delta\sum_{\bm{k}}\left[c^{\dagger}_{+\uparrow}(\bm{k})c^{\dagger}_{+\downarrow}(-\bm{k})-c^{\dagger}_{-\uparrow}(\bm{k})c^{\dagger}_{-\downarrow}(-\bm{k})\right]\nonumber\\
&+{\rm H.c.}\label{HintBCSprime}
\end{align}
differs from $H_{\rm BCS}$ in Eq.\ \eqref{HintBCS} by a $\pi$ phase shift of the pair potential on the top and bottom surfaces. The corresponding pair potential in the BdG Hamiltonian \eqref{HSMB} is 
\begin{equation}
\bm{\Delta}'=\Delta_0\nu_x\tau_z\sigma_0.\label{HSMBprime}
\end{equation}
It anticommutes with $\tau_x$ and thus changes sign upon inversion, representing a pseudoscalar pairing in the classification of Ref.\ \onlinecite{Far16}. 

Bednik, Zyuzin, and Burkov\cite{Bed15} obtain the pseudoscalar pairing \eqref{HintBCSprime} in a model where the pairing interaction is intrinsic to the Weyl semimetal, rather than proximity-induced as in the multilayer structure of Fig.\ \ref{fig_heterostructure}. (The $\tau$ degree of freedom then refers to a molecular orbital instead of to a heterostructure layer.)

The change from scalar to pseudoscalar pairing has drastic consequences for Andreev reflection: The transformed pair potential in Eq.\ \eqref{HtransformedMB},
\begin{equation}
\tilde{\bm{\Delta}}'\equiv{}{\cal V}\bm{\Delta}'{\cal V}^\dagger=-\Delta_0\nu_x\tau_0\sigma_0,\label{Deltatransformedprime}
\end{equation}
is \textit{diagonal} rather than off-diagonal in the $\tau$ degree of freedom. We can therefore project the transformed Hamiltonian,
\begin{align}
\tilde{\cal H}'(\bm{k})={}&v_{\rm F}\nu_z\tau_z(\sigma_x  k_x+\sigma_y  k_y)+M_{k}\nu_0\tau_z\sigma_z+\beta\nu_0\tau_0\sigma_z\nonumber\\
&-\mu\nu_z\tau_0\sigma_0-\Delta_0\nu_x\tau_0\sigma_0,\label{Htransformedprime}
\end{align}
onto the $\tau=-1$ subband without losing the pair potential. There is now no chirality blockade. The Andreev reflection amplitude is
\begin{subequations}
\label{rheXZ0}
\begin{align}
r_{he}={}&-E/\Delta_0+i\sqrt{1-E^2/\Delta_0^2},\\
&\text{at normal incidence for any energy},\nonumber\\
r_{he}={}&\frac{ik_x}{\sqrt{k_x^2+k_y^2}},\\
&\text{at the Fermi level for any angle of incidence}.\nonumber
\end{align}
\end{subequations}

The projected Hamiltonian,
\begin{align}
\tilde{\cal H}'_{\tau=-1}={}&-v_{\rm F}\nu_z(\sigma_x  k_x+\sigma_y  k_y)+(\beta-M_{k})\nu_0\sigma_z\nonumber\\
&-\mu\nu_z\sigma_0-\Delta_0\nu_x\sigma_0,\label{Htransformedprimeproj}
\end{align}
is essentially the one studied in Refs.\ \onlinecite{Che13,Kha16,Mad17}. This explains why no chirality blockade was obtained in those studies of Andreev reflection in a Weyl semimetal.

\subsection{Comparison with tight-binding model simulations}
\label{simulations}

To test these analytical formulas, we have discretized the eight-orbital Hamiltonian \eqref{HSMB} on a cubic lattice, and we solved the scattering problem at the NS interface numerically, using the Kwant toolbox.\cite{kwant} 

Equation\ \eqref{HSMB} is linear in $k_x$ and $k_y$, and a straightforward discretization, by replacing $k_x\mapsto \sin k_x$, $k_y\mapsto \sin k_y$, would suffer from fermion doubling. To avoid this, we follow Ref.\ \onlinecite{Vaz13} and add quadratic terms in $k_x$ and $k_y$ to the mass term $m_k$, resulting in the tight-binding Hamiltonian
\begin{subequations}
\label{HSMBdiscretized}
\begin{align}
&{\cal H}(\bm{k})=\nu_z\tau_z(-\sigma_y  \sin k_x +\sigma_x   \sin k_y )+\beta\nu_0\tau_0\sigma_z\nonumber\\
&\quad+\nu_z(m_{\bm k}\tau_x  -\tau_y  \sin k_z  )\sigma_0-\mu\nu_z\tau_0\sigma_0+\bm{\Delta},\\
&m_{\bm k}=3+\cos k_z -\cos k_x-\cos k_y.
\end{align}
\end{subequations}
For simplicity we have set the Fermi velocity $v_{\rm F}$ and the hopping energies $t_z,t'_z$ equal to unity, and we have taken the same lattice constant $d=a\equiv 1$ parallel and perpendicular to the layers.

The Weyl points are at $\bm{k}=(0,0,\pi\pm K)$, where
\begin{equation}
\begin{split}
&(1-\cos K)^2+\sin^2 K=\beta^2\\
&\Rightarrow K={\rm arctan}\,\left(\frac{\beta\sqrt{4-\beta^2}}{2-\beta^2}\right).
\end{split}
\end{equation}
Near the Weyl point, the normal-state dispersion is
\begin{equation}
\begin{split}
&(E+\mu_{\rm W})^2=k_x^2+k_y^2+q_z^2,\\
&q_z=(\pi- K-k_z)\cos\theta,\;\;\cos^2\theta=1-\beta^2/4.
\end{split}
\end{equation}

\begin{figure}[tb]
\centerline{\includegraphics[width=0.8\linewidth]{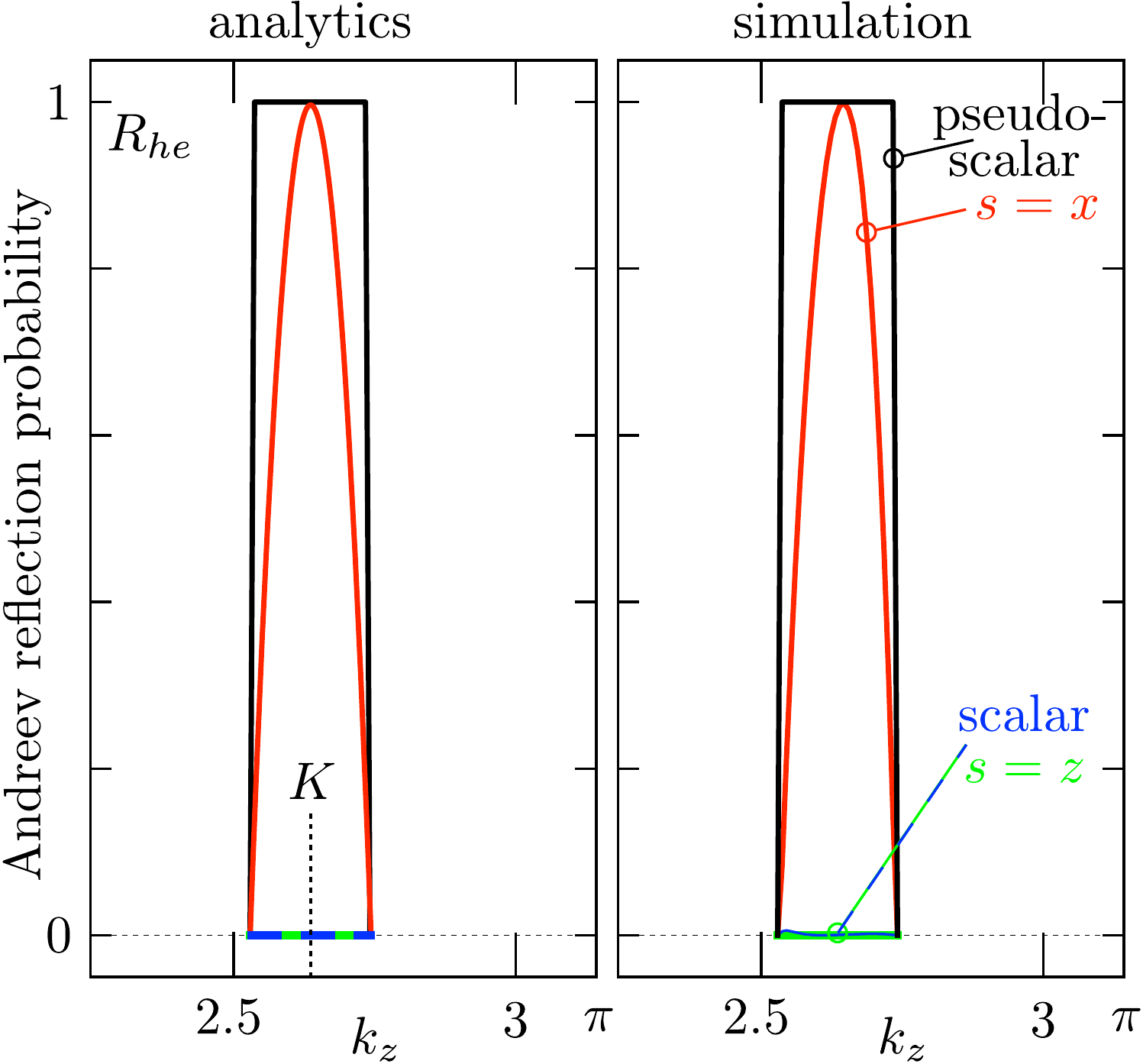}}
\caption{Andreev reflection probability at the Fermi level of a Weyl semimetal -- Weyl superconductor interface, for four different pairing symmetries in the superconductor (scalar and pseudoscalar spin-singlet, and $s=x$ or $s=z$ spin-triplet). The left panel shows the analytical results \eqref{Rheanalytics} for $k_y=0$, $\beta=0.5$, $\mu_{\rm W}=0.1$. The right panel shows the results from a numerical simulation of the tight-binding model \eqref{HSMBdiscretized}, with additional parameters $\mu_{\rm S}=0.4$, $\Delta_0=0.55$. There are two Weyl points at $k_z=\pi\pm K$, only one of which is shown (the other gives the same results).
}
\label{fig_numerics}
\end{figure}

The analytical results for the Andreev reflection probability at the Fermi level ($E=0$), as a function of the transverse momenta $k_y$ and $q_z$, are:
\begin{widetext}
\begin{subequations}
\label{Rheanalytics}
\begin{align}
&R_{he}=\frac{4-\beta^2}{ (2-\beta^2/4)^2}\frac{\mu_{\rm W}^2-k_y^2-q_z^2}{\mu_{\rm W}^2-k_y^2},
&&s=x\;\; \text{triplet pairing},\;\;\bm{\Delta}=\Delta_0\nu_x\tau_y\sigma_x,\\
&R_{he}=0,
&&s=z\;\; \text{triplet pairing},\;\;\bm{\Delta}=\Delta_0\nu_x\tau_y\sigma_z,\\
&R_{he}=\frac{\mu_{\rm W}^2-k_y^2-q_z^2}{\mu_{\rm W}^2-q_z^2},
&&\text{pseudoscalar singlet pairing},\;\;\bm{\Delta}=\Delta_0\nu_x\tau_z\sigma_0,\\
&R_{he}=0,
&&\text{scalar singlet pairing},\;\;\bm{\Delta}=\Delta_0\nu_x\tau_0\sigma_0.
\end{align}
\end{subequations}
\end{widetext}
In Fig.\ \ref{fig_numerics} we compare the analytics with the numerical simulation, and find good agreement without any fit parameter.

\begin{figure}[tb]
\centerline{\includegraphics[width=0.8\linewidth]{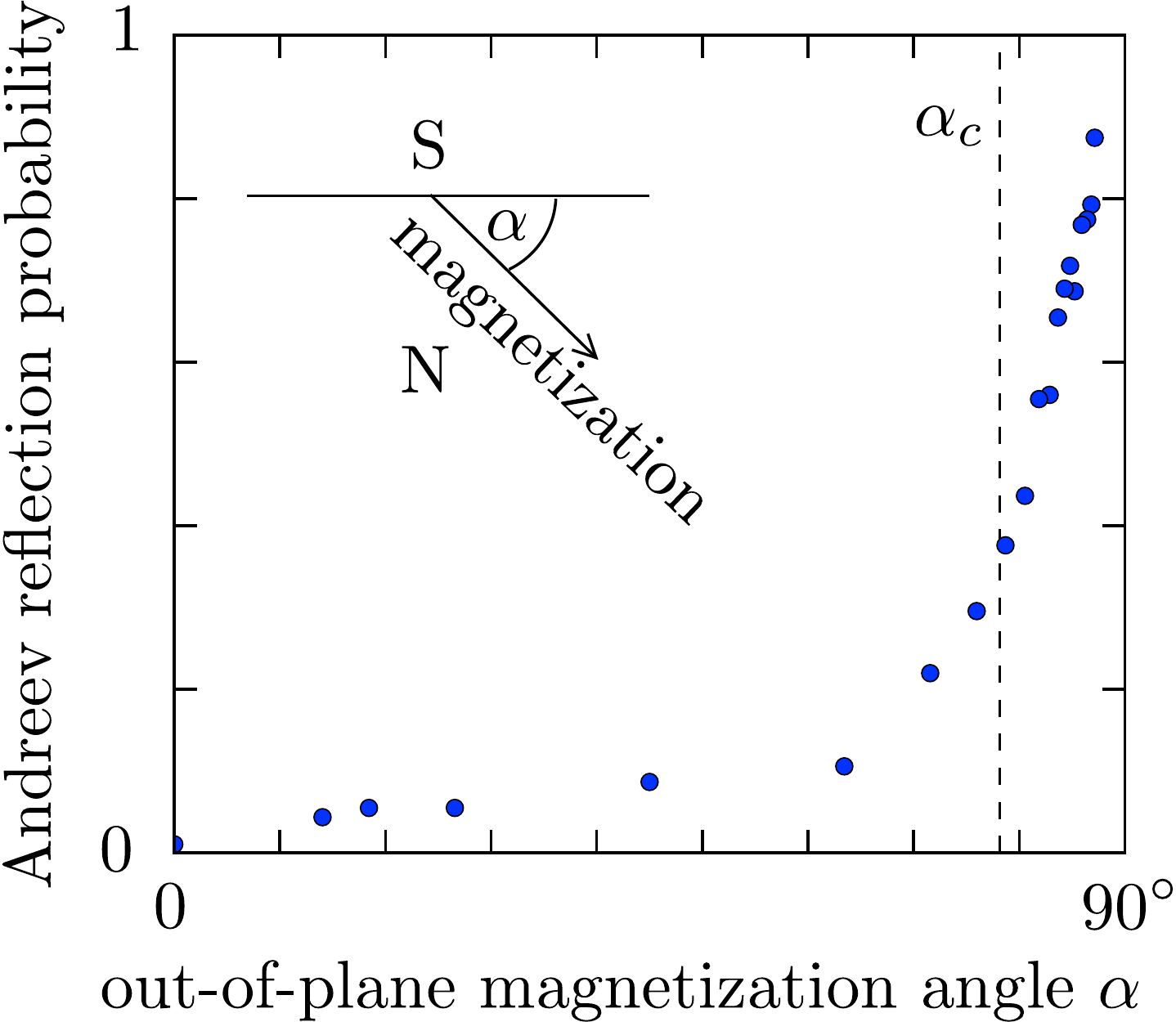}}
\caption{Threshold dependence of the chirality blockade on the direction of the magnetization. The horizontal axis shows the out-of-plane rotation angle $\alpha$ of the magnetization, the vertical axis shows the Andreev reflection probability at the Fermi level for normal incidence. The data points are calculated numerically from the tight-binding model \eqref{HSMBdiscretized} with a scalar pair potential (parameters $\mu_{\rm W}=0.18$, $\mu_{\rm S}=0.2$, $\Delta_0=0.9$, $\beta=0.85$). The dashed vertical line is the threshold angle $\alpha_c={\rm arccos}\,(\mu_{\rm W}/\beta)=78^\circ$ expected for an isotropic Weyl cone.
}
\label{fig_numericsalpha}
\end{figure}

All of this is for a magnetization in the plane of the NS interface. If the magnetization is rotated out of the plane by an angle $\alpha$, the Andreev reflection probability for scalar pairing shows the threshold behavior discussed in Sec.\ \ref{sec_model}, see Fig.\ \ref{fig_numericsalpha}. The threshold angle given by $\cos\alpha_{\rm c}=k_{\rm F}/K\approx\mu_{\rm W}/\beta$ for an isotropic Weyl cone is in reasonable approximation with the numerical result, with some deviations because the Weyl cone of the Hamiltonian \eqref{HSMBdiscretized} has a significant anisotropy.

\section{Calculation of the Fermi-arc mediated Josephson effect}
\label{FermiarcJcalc}

We calculate the supercurrent flowing through an SNS junction in response to a phase difference $\phi$ between the superconducting pair potentials. As explained in Sec.\ \ref{FermiJosephson}, because of the chirality blockade of Andreev reflection this supercurrent is due entirely to overlapping Fermi arcs on the two NS surfaces. It is exponentially small when the distance $L$ of the NS interfaces is large compared to the decay length $v_{\rm F}/\beta$ of the surface states into the bulk. This is the key difference between the present calculation for the Weyl semimetal Josephson junction and a similar calculation for a graphene Josephson junction in Ref.\ \onlinecite{Tit06}.

\subsection{Andreev bound states}

We start from the Hamiltonian \eqref{HSMB},
\begin{align}
{\cal H}={}&v_{\rm F}\nu_z\tau_z(-\sigma_y   k_x +\sigma_x   k_y )+\beta\nu_0\tau_0\sigma_z\nonumber\\
&+\nu_z(m_k\tau_x  -\tau_y  t_z\sin k_z d )\sigma_0-\mu\nu_z\tau_0\sigma_0\nonumber\\
&+\Delta_0(\nu_x\cos\varphi-\nu_y\sin\varphi)\tau_0\sigma_0,\label{HSMBapp}
\end{align}
generalized to account for a complex pair potential $\Delta_0 e^{i\varphi}$. In the N region $|x|<L/2$ we set $\Delta_0=0$, while in the S regions $|x|>L/2$ we have a nonzero gap $\Delta_0$ and a phase $\varphi$ equal to $\phi/2$ for $x>L/2$ and equal to $-\phi/2$ for $x<-L/2$.

We carry out a (partial) block-diagonalization by means of the unitary transformations ${\cal H}\mapsto{\cal W}{\cal V}{\cal H}{\cal V}^\dagger{\cal W}^\dagger$, with ${\cal V}$ defined in Eq.\ \eqref{UBBdef} and ${\cal W}$ defined by
\begin{equation}
{\cal W}=\begin{pmatrix}
i\tau_y\sigma_z&0\\
0&\tau_0\sigma_0
\end{pmatrix}.\label{calWdef}
\end{equation}
The resulting Hamiltonian
\begin{align}
{\cal H} ={}& v_{\rm F}\nu_z\tau_z\big(k_x\sigma_x+k_y \sigma_y - M_k \sigma_z\big)
+\beta\nu_0\tau_0\sigma_z \nonumber \\
& -\mu \nu_z\tau_0\sigma_0+\Delta_0(\nu_x\cos\varphi-
\nu_y\sin\varphi)\tau_0\sigma_0\label{calHWVtransformed}
\end{align}
is diagonal in $\tau$. We may therefore replace $\tau_z$ by the variable $\tau=\pm1$ and $\tau_0$ by $1$.

At the NS interfaces $x=\pm L/2$ we have the boundary condition \eqref{psihpsiechi2}, 
\begin{equation}
\begin{split}
&\psi_h(\pm L/2) = {\cal T}^{\pm 1}\,\psi_e(\pm L/2), \\
&{\cal T} = e^{-i\phi/2}e^{-i\alpha \tau \sigma_x},\;\;\;\;
\alpha= {\rm arccos} (E/\Delta_0 ).
\end{split}
\end{equation}
Integration of ${\cal H}\psi=E\psi$, with $\psi=(\psi_+,\psi_-)$ the two $\nu$-components of the wave function, gives the $x$-dependence in the N region,
\begin{subequations}
\begin{align}
&\psi_{\pm}(x)=e^{x\,\Xi_{\pm}}\psi_{\pm}(0),\;\;-L/2<x<L/2,\\
&\Xi_{\pm}=i\tau\sigma_x\frac{\mu\pm E}{v_{\rm F}}
-\sigma_y\frac{ M_k\pm\tau\beta}{v_{\rm F}} +\sigma_z k_y.
\end{align}
\label{Xi}
\end{subequations}

A bound state in the SNS junction, a socalled Andreev level, appears at energies when\cite{Tit06}
\begin{equation}
\det\big(1-e^{-L\,\Xi_+}\,{\cal T}\,e^{L\,\Xi_-}\,{\cal T} \big)=0.
\end{equation}
We assume that the separation $L$ of the NS interfaces is small compared to the superconducting coherence length $\xi=v_{\rm F}/\Delta_0$. In this short-junction regime the energy dependence of $\Xi_\pm$ can be neglected and only the energy dependence of ${\cal T}$ needs to be retained.\cite{Bee91}

Introducing the vector notation 
$\boldsymbol{\sigma}=\big(\sigma_x,\sigma_y,\sigma_z\big)$
and defining
\begin{align}
&\mathbf{d}_\pm=\big(d_x,d_{\pm, y},d_z\big)=\frac{L}{v_{\rm F}}\big(i\tau\mu,
-M_k\pm\tau\beta,v_{\rm F}\sin k_y\big),
\label{ds}
\end{align}
the bound-state condition can be written as
\begin{align}
&\det\big(e^{-i\phi/2}e^{\mathbf{d}_-\cdot\boldsymbol{\sigma} }-e^{i\phi/2}\,e^{i\alpha\sigma_x}
\,e^{\mathbf{d}_+\cdot\boldsymbol{\sigma}}\,e^{i\alpha\sigma_x} \big)=0.
\label{cbs}
\end{align}

To simplify the equations we define
\begin{equation}
{\rm sinhc}\,x=\frac{\sinh x}{x}.\label{sinhcdef}
\end{equation}
The identity
\begin{align}
e^{\mathbf{d}\cdot\boldsymbol{\sigma} } =&{} \sigma_0\cosh d+(\mathbf{d}\cdot\boldsymbol{\sigma})\,{\rm sinhc}\, d,
\;\;\;\; d=\sqrt{\mathbf{d}\cdot\mathbf{d}},
\end{align}
allows us to evaluate the determinant Eq.\ \eqref{cbs} as
\begin{equation}
\gamma_0^2=\gamma_1^2+\gamma_2^2+\gamma_3^2,\label{gammaequation}
\end{equation}
where
\begin{subequations}
\begin{align}
\gamma_0 ={}& 
e^{-i\phi/2}\cosh d_- 
-e^{i\phi/2}\bigl(\cos2\alpha\,\cosh d_+ \nonumber \\
& +i\sin 2\alpha\,
d_x\,{\rm sinhc}\,d_+\bigr),\\
\gamma_1 ={}& 
e^{-i\phi/2}d_x\,{\rm sinhc}\,d_- -e^{i\phi/2}
\bigl(i\sin 2\alpha\,\cosh d_+ \nonumber\\ 
&+\cos 2\alpha\,
d_x\,{\rm sinhc}\,d_+\bigr),\\
\gamma_2 ={}& 
e^{-i\phi/2}d_{-,y}\,{\rm sinhc}\,d_- -e^{i\phi/2}
d_{+,y}\,{\rm sinhc}\,d_+, \\
\gamma_3 ={}& 
e^{-i\phi/2}d_{z}\,{\rm sinhc}\,d_- -e^{i\phi/2}
d_{z}\,{\rm sinhc}\,d_+ .
\end{align}
\end{subequations}

The phase dependence of the bound-state energy can be solved exactly from Eq.\ \eqref{gammaequation} when the Fermi level is near the Weyl points, $|\mu|\ll v_{\rm F}/L$:
\begin{subequations}
 \label{ephi}
 \begin{align}
E(\phi) ={}& \Delta_0\sqrt{\tfrac{1}{2}+p(\phi)} \\
p(\phi) ={}& \frac{1+(\mathbf{d}_-\cdot\mathbf{d}_+)\,{\rm sinhc}\, d_-\,{\rm sinhc}\, d_+
}{2\cosh d_-\cosh d_+} \nonumber\\
&-  \frac{\sin^2(\phi/2)}{\cosh d_-\cosh d_+},
\end{align}
\end{subequations}
where the $\mathbf{d}_\pm$ are taken at $\mu=0$. The energy levels are doubly degenerate in 
$\tau=\pm1$. This degeneracy is 
lifted by a finite chemical potential: The first-order correction $\delta E_\pm$ to the bound state energy reads
\begin{equation}
\delta E_\pm= \pm\frac{L|\mu|\Delta_0}{2v_{\rm F}}\,
\Big|\frac{\tanh d_+}{d_+}-\frac{\tanh d_-}{d_-} \Big|\sqrt{\tfrac{1}{2}-p(\phi)}.
\end{equation}

\subsection{Josephson current }

In the short-junction limit only the Andreev levels contribute to the supercurrent density,\cite{Bee91} according to
 \begin{equation}
 j(\phi)=-\frac{e}{\hbar} \sum_{\tau=\pm1}\int_{-\pi}^\pi\frac{dk_y}{2\pi}\int_{-\pi}^\pi\frac{dk_z}{2\pi}\, \frac{dE(\phi)}{d\phi}.
 \end{equation}
We take $\mu\ll v_{\rm F}/L$ and substitute Eq.\ \eqref{ephi}, to arrive at
\begin{equation}
j(\phi) = \frac{e\Delta_0}{8\pi^2\hbar}
\int_{-\pi}^\pi dk_y\int_{-\pi}^\pi d k_z \frac{\sin\phi}{ \cosh d_-\cosh d_+
\sqrt{\tfrac{1}{2}+p(\phi)}}.\label{jphiresult}
\end{equation}
(The integrand is symmetric in $\tau=\pm$, so the sum over $\tau$ has been omitted in favor of an overall factor of $2$.)

\begin{figure}
\includegraphics[width=1\linewidth]{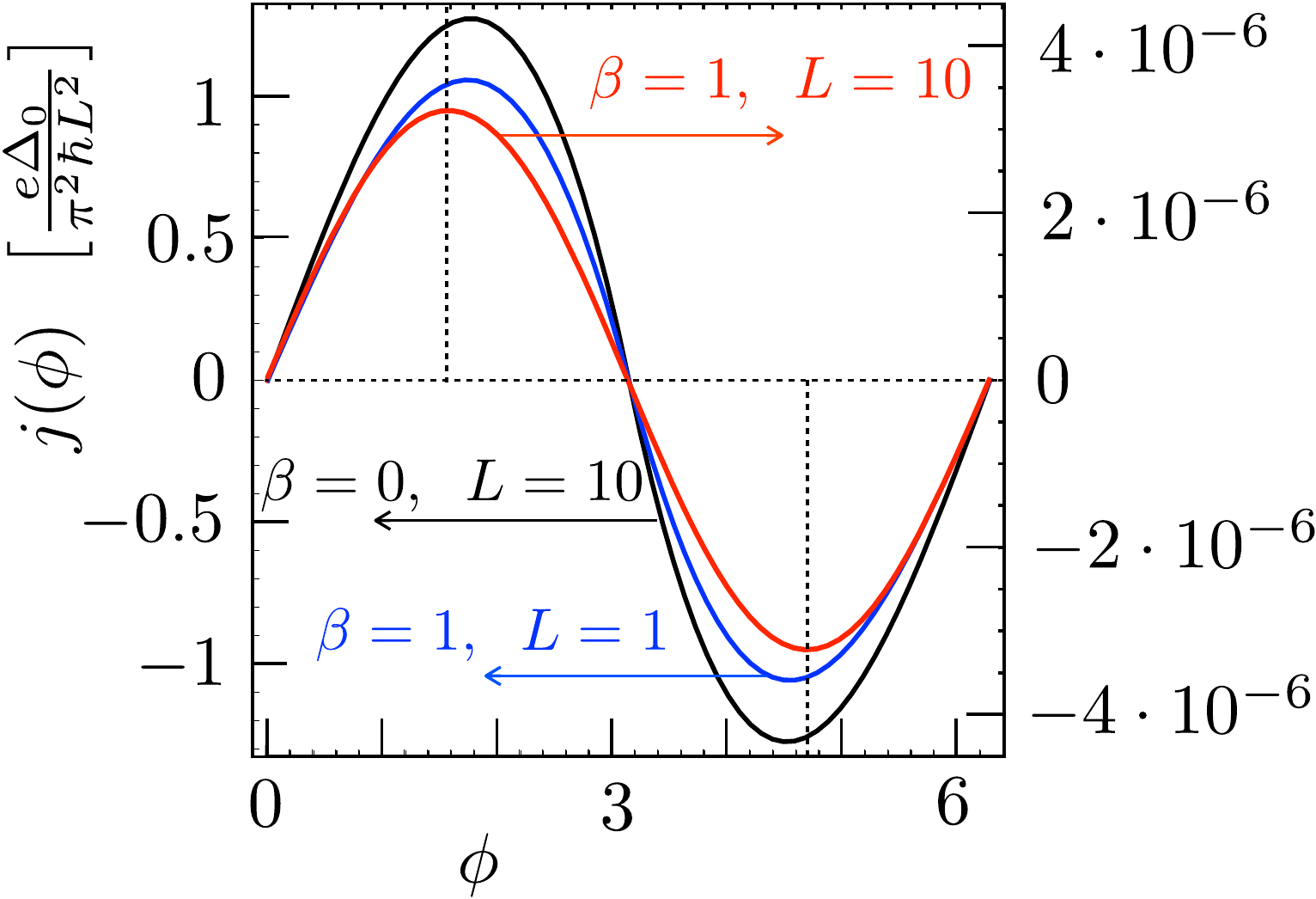}
\caption{Current-phase relationship of the Josephson current density for various values of 
$\beta$ and $L$. The extrema are close to $\pi/2$ and $3\pi/2$, indicated
by the dashed lines. This is calculated from Eq.\ \eqref{jphiresult} for $v_{\rm F}=t_z=t_z'=d=1$.} 
\label{current-phase}
\end{figure}

We take parameters $v_{\rm F}=t_z=t_z'=d=1$, when
\begin{equation}
M_{k}=\sqrt{(1+\cos k_z)^2+\sin^2 k_z }=2\cos (k_z/2).
\end{equation}
The current-phase relationship is close to sinusoidal, see Fig.\ \ref{current-phase}. The critical current can then be accurately approximated by $j_c\approx j(\pi/2)$. This is plotted as a function of $L$ in Fig.\ \ref{Jcurrent}. It decays $\propto \exp(-L/\xi_{\rm arc})$, with $\xi_{\rm arc}\simeq  v_{\rm F}/\beta$ the penetration depth of the surface Fermi arc into the bulk.

\begin{thebibliography}{99}
\bibitem{note0} A quick way to see that electrons and holes in a given spin band and Weyl cone move in the same direction, is to note that their dispersion relations are related by $E_{\rm electron}(k)=-E_{\rm hole}(-k)$, so a linear dispersion $E=v_{\rm F}k$ remains the same.
\bibitem{note1} It is justified to ignore electron-hole coupling in the interface matrix \eqref{MSWdef}, because this describes the transmission through a length of order $1/k_{\rm F}$ in S, which is smaller than the electron-hole coupling length $v_{\rm F}/\Delta_0$ by a factor $E_{\rm F}/\Delta_0\gg 1$.
\bibitem{MBnote} We should note that our Eq.\ \eqref{HtransformedMB}, with a pair potential that is off-diagonal in $\tau$, disagrees with Eqs.\ 5--8 of Ref.\ \onlinecite{Men12}, which have a fully block-diagonal Hamiltonian in the layer degree of freedom. We do agree on the Hamiltonian before the unitary transformation [our Eq.\ \eqref{HSMB} and Eqs.\ 1--3 in Ref.\ \onlinecite{Men12}] and we have traced the discrepancy to an algebraic error.
\bibitem{twoband} The Hamiltonian $H'_{\rm W}=(H_+,H'_-)$ is called a two-band model because near the Weyl points Eqs.\ \eqref{Hplusdef} and \eqref{Hminprimedef} can be combined into a single $2\times 2$ matrix of the form $v_x k_x\sigma_x+v_y k_y\sigma_y + (v_z/2K) (k_z^2-K^2)\sigma_z$. No additional orbital degree of freedom is needed, as in the four-band model \eqref{HW}. The inversion-symmetry breaking remains hidden unless there are spin-dependent processes that couple the Weyl cones, as in Andreev reflection.
\end{thebibliography}
\end{document}